\documentclass[11pt, onecolumn, draftcls]{IEEEtran}

\usepackage{amsmath,graphicx}
\usepackage{subfigure}
\usepackage{epstopdf}
\usepackage{float}
\usepackage{setspace}
\usepackage{slashbox}
\usepackage{url}
\usepackage{multirow}
\usepackage{pstricks}
\usepackage{pst-sigsys}
\usepackage{pstricks-add}
\usepackage[crop=off]{auto-pst-pdf}
\usepackage{pst-node}
\usepackage{threeparttable}
\usepackage{cite}

\graphicspath{{image_dir/}{pdf_dir/}}

\begin{document}
\doublespacing

\title{Invisible Flow Watermarks for Channels with Dependent Substitution, Deletion,  and Bursty Insertion Errors
(Draft)}

\author{ Xun~Gong,~\IEEEmembership{Student~Member,~IEEE},
        Mavis~Rodrigues, 
        Negar~Kiyavash,~\IEEEmembership{Member,~IEEE}
        \thanks{This work was supported in part by AFOSR under Grant FA9550-11-1-0016,  MURI under AFOSR Grant FA9550-10-1-0573, and NSF CCF 10-54937 CAR. This work was presented in part at ICASSP'12~\cite{xunICASSP12}. 
 
X. Gong is with the Coordinated Science Laboratory and the Department of Electrical Engineering, University of Illinois at Urbana-Champaign, Urbana, IL 61801 USA (email: {xungong1@illinois.edu})

M. Rodrigues was with the Coordinated Science Laboratory and the Department of Electrical Engineering, University of Illinois at Urbana-Champaign, Urbana, IL 61801 USA (email:{mavis.rodrigues@gmail.com})
 
 N. Kiyavash is with  the Coordinated Science Laboratory  and the Department of Industrial and Enterprise Systems Engineering, University of Illinois at Urbana-Champaign, Urbana, IL 61801 USA (email: {kiyavash@illinois.edu})        
}
}
\maketitle

\begin{abstract}
Flow watermarks efficiently link packet flows in a network in order to thwart various attacks such as stepping stones.
We study the problem of designing good flow watermarks. 
Earlier flow watermarking schemes mostly considered substitution errors, neglecting the effects of packet insertions and deletions that commonly happen within a network. More recent schemes consider packet deletions but often at the expense of the watermark visibility. 
We present an invisible flow watermarking scheme capable of enduring a large number of packet losses and insertions.  To maintain invisibility, our scheme uses quantization index modulation (QIM) to embed the watermark into inter-packet delays, as opposed to time intervals including many packets. As the watermark is injected within individual packets, packet losses and insertions may lead to watermark desynchronization and substitution errors.
To address this issue, we add a layer of error-correction coding to our scheme. Experimental results on both synthetic and real network traces demonstrate that our scheme is robust to network jitter, packet drops and splits, while remaining invisible to an attacker.
\end{abstract}

\IEEEpeerreviewmaketitle

\section{Introduction}
\label{sec:intro}
\PARstart{D}etecting correlated network flows, also known as flow linking, is a  technique for traffic analysis with wide applications in network security and privacy.
For instance, it may be utilized to expose a stepping stone attacker who hides behind proxy hosts.
Alternatively, flow linking has been successfully used to attack low-latency anonymity networks such as Tor~\cite{Wang07}, where anonymity is compromised once end flows are correctly matched. 
As network connections are often encrypted, it is infeasible to link flows directly relying on packet  contents.
However, matching flows using side information such as packet timings is possible, as their values remain to some extent unchanged even after encryption~\cite{Staniford-Chen95, Zhang00, Yoda00}.

Earlier work in flow linking was  based on long observation of flow characteristics, such as the number of ON/OF periods~\cite{Zhang00}.
Such {\em passive} techniques are fragile vis-a-vis network artifacts and require long observation periods to avoid large false alarm rates.
{\em Flow watermarking}, an active approach, was suggested as an improvement. In this approach,  a pattern, the watermark, is injected into the flow with the hope that the flow stays traceable after traversing the network as long as the same pattern can be later extracted~\cite{Wang02, Wang05, Wang07, Pyun07,Houmansadr09,Yu07}. 
Unlike passive schemes, flow watermarking is highly reliable and  works effectively on short flows. 

The challenge of designing good flow watermarks is to keep the injected pattern robust to network artifacts yet invisible to watermark attackers.\footnote{The goal of watermark attackers is to prevent the success of flow linking by disrupting the detection or altogether removing the watermarks from the flow.}
The robustness requirement guarantees that the injected pattern survives network artifacts, while
the invisibility property prevents  watermark removal attempts by active attackers.
Most state-of-the-art schemes currently trade off one of the two properties at the expense of the other. 
In the so called \emph{interval-based} schemes~\cite{Wang07, Pyun07}, a flow is divided into intervals, and all packets within selected intervals are shifted to form a watermark pattern.
Given that  a few packets would not greatly affect the pattern created in the entire interval, these schemes are robust against network artifacts such as packet drops and splits.
However, shifting a large number of packets produces noticeable `traces'  of the embedded watermarks and compromises the invisibility requirement~\cite{Kiyavash08}. 
In \emph{inter-packet-delay (IPD)-based} schemes~\cite{Wang02,Houmansadr09}, the delays between consecutive packets are modulated to embed watermarks.
Since only small perturbations are introduced in each inter-arrival time,  watermarks are not visible. 
The drawback of this approach is that  any packet loss or insertion during transmission  can cause watermark desynchronization and severe  decoding errors.

In this paper,  we present a new IPD-based flow watermarking scheme where invisible watermark patterns are injected in the inter-arrival-time of  successive packets.
We treat the network as a channel with substitution, deletion, and bursty insertion errors caused by jitter, packet drops, and packet splits or retransmission, respectively, and introduce an insertion, deletion and substitution (IDS) error-correction coding scheme to communicate the watermark reliably over the channel. 
At the same time,  we preserve watermark invisibility  by making unnoticeable  modifications to packet timings using the QIM framework~\cite{Chen01}. 
Through experiments on both synthetic and real network traces, we show that our scheme performs reliably in presence of network jitter, packet losses and insertions.
 Furthermore, we verify the watermark invisibility using {\em Kolmogorov-Smirnov} ~\cite{Massey51} and {\em multi-flow-attack} tests~\cite{Kiyavash08}.
 Deletion correction codes were first applied to flow watermarking in~\cite{xunICASSP12}, where watermarks can be decoded correctly after packet losses as long as the first packet in the flow was not dropped. In this work, we extend our decoder to  handle more realistic network environments where not only packet losses but also packet insertions occur. 
Furthermore, synchronization requirement on the first packet is relaxed. 
To verify the performance of our scheme,  traffic traces collected from real SSH connections  are tested. This improves the simulation setup in~\cite{xunICASSP12}, where merely  simulated synthetic traffic was used.

The rest of the paper is organized as follows. 
Background  on flow watermarking appears in \S\ref{sec:background}. We describe notations and definitions in \S\ref{sec:def}. Our proposed scheme is presented in \S\ref{sec:scheme}.
We evaluate the performance of our scheme using synthetic and real traffic traces in \S\ref{sec:eval}. 

\section{Background}
\label{sec:background}
This section covers some background material on flow watermarking. First, we describe three application scenarios of flow watermarking. Second,  we discuss some  principles for designing good watermarking schemes. 
We conclude by surveying the literature.

\subsection{Applications}
\begin{figure}[t]  
 \centering
   \includegraphics[width=0.7\columnwidth]{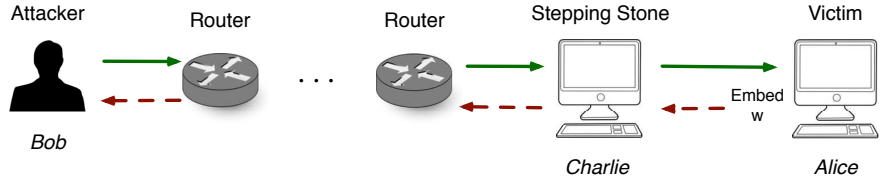}
   \caption{Detecting attackers behind stepping stones.  {\em Bob} uses {\em Charlie} as a stepping stone to attack {\em Alice} so that his identity remains hidden from {\em Alice}.  To traceback the origin of this attack, {\em Alice} injects a watermark on the flow sent back to the stepping stone. The path leading to {\em Bob} is exposed as every router along this path detects {\em Alice}'s watermark on  flows passing through.}
   \label{fig:step_stone}
\end{figure}

We begin with a stepping-stone detection scenario where flow watermarks are used to find hidden network attackers.  
Figure~\ref{fig:step_stone} depicts an attacker {\em Bob} who wants to  attack a victim {\em Alice} without exposing his  identity. 
{\em Bob} first remotely logins to a compromised intermediate host {\em Charlie} via SSH~\cite{ssh}.
Then he proceeds by sending  attack flows to {\em Alice} from {\em Charlie}'s machine.
Tracing packet flows sent to Alice's machine would implicate {\em Charlie} instead of {\em Bob} as the attacker.
Hosts like  {\em Charlie}, exploited to hide the real attack source, are called as {\em stepping stones}~\cite{Staniford-Chen95}. 
In real life, attackers may hide behind a chain of stepping stones, making it hard for the victim, who only sees the last hop, to determine the origin of the attack.
Fortunately, flow watermarking is a solution for tracing the attack source. 
Notice that an interactive connection is maintained along {\em Bob-Charlie-Alice} during the above stepping stone attack.
Hence  {\em Alice} can secretly embed a watermark in the packet flow heading back to {\em Charlie}. As this flow travels back to {\em Bob}, the watermark could be subsequently detected by the intermediate routers (or firewalls), revealing the attack path and its true origin~\cite{Yung02, Ding11}. 

\begin{figure*}[t]
 \centering
 \subfigure[]{
  \includegraphics[width=0.45\columnwidth]{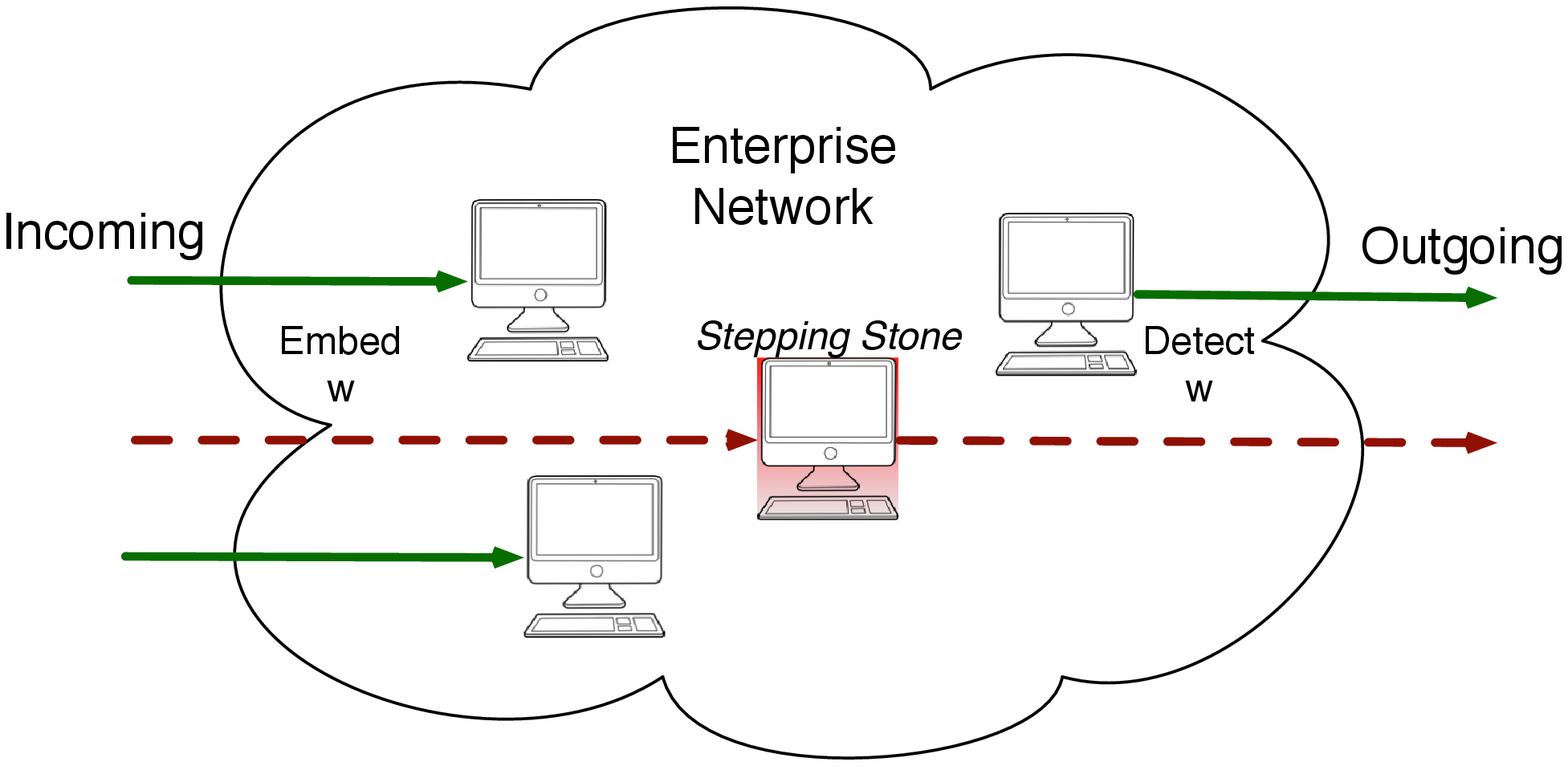}
   \label{fig:step_stone_enterprise}
}\subfigure[]{
  \includegraphics[width=0.45\columnwidth]{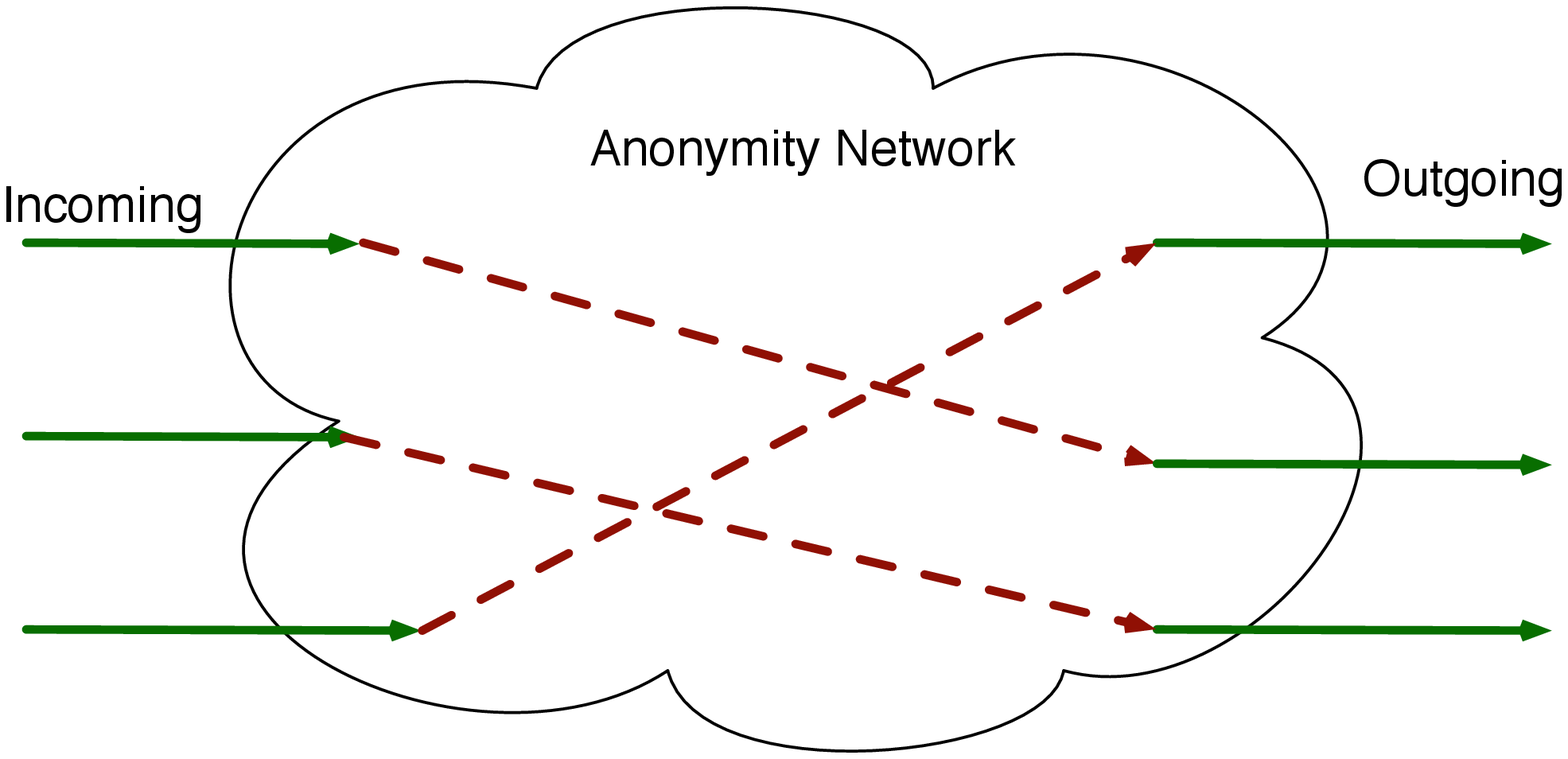}
   \label{fig:anonymity}
   }
   \caption{(a) Stepping stones in enterprise networks. An intruder compromises a host in the enterprise network as a `stepping stone'. The enterprise embeds watermarks on all incoming flows and monitors all the outgoing flows. Any pair of incoming/outgoing flows with the same watermark indicates the existence of inside stepping stones. (b) An anonymity network. Incoming flows are shuffled before leaving the system to hide the pairing among communicating parties.}
 \label{fig:application}
\end{figure*}

Another scenario of stepping-stone attack occurs in enterprise networks, as shown in Figure~\ref{fig:step_stone_enterprise}. Here, intruders are trying to compromise hosts in an enterprise network to relay their malicious traffic~\cite{Lippmann05, Kiyavash08}. 
To discover this kind of `stepping stones' within the network, an enterprise can add watermarks on all incoming flows, and then terminate outgoing flows that contain the watermark since they most probably come from stepping stones.
In a similar fashion, flow watermarking may be applied to attacking anonymity network systems~\cite{ssh,i2p,freenet,gnunet}.
In order to hide the identities of communicating parties, an anonymity network shuffles all the flows passing through it, as shown in Figure~\ref{fig:anonymity}. 
If an attacker somehow discovers the hidden mappings between incoming and outgoing flows, the anonymity is compromised. 
Akin to the previous enterprise network scenario, this can be achieved by marking all incoming flows with watermarks and subsequently detecting the watermarks on the exiting flows. 

\subsection{Design Principles}
\label{sec:design}

From above application examples, we summarize a list of principles for designing flow watermarks.
The challenge of building an efficient scheme lies in the difficulty of achieving all desired properties simultaneously.  

\begin{itemize}
\item {\em Robustness}.
One major advantage of flow watermarking over passive traffic analysis is the robustness against network noise. 
Take the stepping stone attack of Figure~\ref{fig:step_stone} for example. 
The flow {\em Alice} sends back to {\em Bob} is subjected to jitter, packet drops, and packet splits during transmission. 
All these artifacts can alter the watermark, resulting in decoding errors. 
Without the ability to withstand these artifacts,  flow watermarking is no different than passive analysis, which is fragile  by nature. 

\item {\em Invisibility}.
A successful watermark pattern should stay `invisible' to avoid possible attacks. 
For instance, in Figure~\ref{fig:step_stone_enterprise},
if the intruder notices  that incoming flows contain watermarks,  it can command the stepping stone to take precautionary actions (for instance, remove the watermarks altogether).

\item {\em Blindness}.
In a blind watermarking scheme, the watermark pattern can be extracted without the help of the original flow~\cite{Cox}. 
On the contrary, the original flow must be present in order to detect non-blind watermarks. Again, consider the example of Figure~\ref{fig:step_stone_enterprise}. 
In order to detect the stepping stone, the enterprise needs to perform watermark decoding on all outgoing flows.
If a non-blind detection scheme is used, all exit routers are required to obtain a copy of each incoming flow. 
The resulting overheads of bandwidth and storage make such schemes impractical  in large enterprise networks.
\item{\em Presense watermarking}.
In conventional digital watermarking (e.g., multimedia watermarking), often a large amount of hiding capacity is desired as the injected watermarks are frequently used to achieve copyright among many users~\cite{Sergio}. This, fortunately, is not required for most flow watermarking applications, since the main purpose of injecting watermarks here is to link flows initiated from the same sources. 
In other words, in digital watermarking terminology,  {\em zero-bit} or {\em presence} watermarks suffice~\cite{Cox}.
Therefore, when designing a flow watermarking scheme, one may trade the capacity for other properties such as robustness (see the discussion in~\S\ref{sec:ids_enc}).
\end{itemize}

\subsection{Watermark Attack Models}
The difficulty of maintaining watermark invisibility depends on the specific attack model. 
Based on the strength of the watermark attacker, attack models may be classified as follows:
\begin{itemize}
\item {\em Level-I:} the attacker observes  the watermarked flow, and has knowledge of certain feature (e.g., empirical distributions of IPDs) of the original flow;

\item {\em Level-II:} the attacker observes  the watermarked flow, and  has a distorted version of the original flow;

\item {\em Level-III:} the attacker observes both the watermarked flow and the original flow. 
\end{itemize} 

In Level-I, the weakest attack model, the attacker can only discover the presence of watermark by statistical approaches that real a deviation of know features from the norm in the original flow. 
 For interval based schemes, the multi-flow attack (MFA) exposes empty intervals in the combination of several watermarked flows~\cite{Kiyavash08}.  
For IPD-based schemes, the empirical distribution of IPDs, which should not be changed with high probability, 
distinguishes watermarked flows from unwatermarked ones via Kolmogorov-Simirnov (K-S) tests~\cite{Massey51,Houmansadr09}. 
We show in~\S\ref{sec:visibility} that our watermark does not introduce noticeable patterns for the MFA or the KS test to detect it. 

In Level-II, given a distorted version of the original flow, the attacker has  in effect an imperfect realization of  the original flow signal which is more informative than the statistical information a Level-I attacker has. 
A Level-II attack, BACKLIT was recently proposed, where the attacker serves as a traffic relay between the client and server of a TCP connection and thus sees both  REQUEST and RESPONSE flows~\cite{Luo}. 
When watermarks are added, packets along one direction (i.e., RESPONSE) must be delayed. The attacker can detect this `delayed' timing pattern as he observes the `clean' flow in the REQUEST direction.
BACKLIT works well when a strong correlation between the REQUEST and RESPONSE flows exists, in which case the attack has a high fidelity version of the original flow.  
In~\S\ref{sec:visibility}, we evaluate our scheme against BACKLIT. We show that in practice the correlation between the response and request flows are destroyed for the most part
 by network jitter because watermarks in our scheme that add very small perturbations to IPDs can remain hidden. 

A Level-III attacker, who observes the exact original flow has a significantly easier task detecting the presence of a watermark~\cite{Lin}. 
This attack model, however,  requires the attacker to be able to observe arbitrary flows everywhere in the network, which is impossible for most real applications. In this work, we focus on the first two attack models when evaluating watermark invisibility. 


\subsection{Related Work}
\label{sec:related}
We briefly review previous flow watermarking literature. 
To the best of our knowledge, all the previous schemes fail to meet  at least one of the above design principles, 
necessitating the development of a comprehensive approach that meets all the aforementioned criteria.

Earlier flow watermarks are of {\em inter packet delay (IPD)-based} type.
In~\cite{Wang02}, the authors propose an IPD-based scheme that modulates the mean of selected IPDs using the QIM framework. 
Watermark synchronization is lost if enough packets are dropped or split.
Therefore the scheme is unreliable. 
Another IPD-based scheme is presented in~\cite{Houmansadr09}, where watermarks are added by enlarging or shrinking the IPDs. 
This non-blind scheme achieves some watermark resynchronizations when packets are dropped or split, but is not scalable as the original packet flow is required during decoding. 

In {\em interval-based} schemes, instead of using the IPDs between individual packets, the watermark pattern is encoded into batch packet characteristics within fixed time intervals. 
In~\cite{Wang07}, an interval-centroid scheme is proposed. 
After dividing the flow into time intervals of the same length, the authors create two patterns by manipulating the centroid of packets within each interval. The modified centroids are not easily changed even after packets are delayed, lost or split. 
A similar design is presented in~\cite{Pyun07}, where the watermark pattern is embedded in the packet densities of predefined time intervals.
One problem with interval-based schemes is the lack of invisibility. 
Moving packets in batches generates visible artifacts, which can expose the watermark positions. 
Based on this observation,   a multi-flow attack (MFA) was proposed in~\cite{Kiyavash08}.
The authors showed that by lining up as few as 10 watermarked flows, an attacker can observe a number of large gaps between packets (see Figure.10 in~\cite{Kiyavash08}) in the aggregate flow,  revealing the watermark  positions.
Recently, a new  interval-based scheme was proposed in~\cite{Swirl}. The main idea is that the exact locations of modified intervals depends on the flow pattern. This flow-dependent design reduces the success rate of MFA, but makes it more difficult to retrieve the correct intervals for decoding in face of strong network noise. 
Moreover, the perturbation introduced in the IPDs is large enough to make the scheme susceptible to Level-II attacks such as BACKLIT. 
\begin{table}
\centering
\caption{Summary of current watermarking schemes}
\begin{tabular}{|c|c|cccc|}
\hline
& & \shortstack{Invisibility\\Level-I} & \shortstack{Invisibility\\Level-II} & Robustness & Blindness\\ \hline
\multirow{3}{*}{Interval-based} & \cite{Wang07} & no & no &  yes & yes\\
& \cite{Pyun07} & no & no & yes & yes\\ & \cite{Swirl} &yes&no&yes&yes\\\hline
\multirow{3}{*}{IPD-based} & \cite{Wang02} & yes & yes$^{*}$  & no  & yes\\
& \cite{Houmansadr09} & yes  & no & yes  & no\\
& The proposed scheme & {\bf yes} & {\bf yes}$^{*}$  & {\bf yes}  & {\bf yes}\\ \hline
\end{tabular}
 \begin{tablenotes}
        \footnotesize
        \item *The Level-II attack model is effective only when network jitter is small. Schemes like ours that add very small perturbations to IPDs remain hidden under normal network operating conditions (See~\S\ref{sec:visibility}).

        \end{tablenotes}
\label{tab:summary}
\end{table}

Table~\ref{tab:summary} compares existing flow watermarking schemes with our proposed scheme. 
Unlike previous work, the new scheme satisfies all the desired properties.

\section{Notations and Definitions}
\label{sec:def}

In the discussion of the rest of the paper, we use the following notation.
$\mathbf{a}^b=\{a_1,a_2,\cdots,a_b\}$ is a  sequence of length $b$; 
$a^{r}_{t}=\{a_t,\cdots,a_r\}$ is a sequence in $\mathbf{a}^b$ starting with index $r$ and ending with $t$. Specially, if $r\leq t$, $a^{r}_{t}$ is an empty sequence, denoted by $\emptyset$; 
$\oplus$ denotes the `xor' operation.

We also define the following variables used in our scheme. 
\begin{itemize}
\item $\mathbf{I}^M$ is the IPD sequence of an original packet flow, where each delay, $I_i$,  is positive real valued;
\item  $\mathbf{I'}^M$ is the IPD sequence of the same flow after injection of the watermark pattern;
\item $\hat{\mathbf{I}}^{M'}$ is the IPD sequence received by the watermark decoder;
\item $\mathbf{w}^n$ is the binary  watermark sequence; 
\item $\mathbf{\tilde{w}}^N$ is a sparse version of $\mathbf{w}^N$, where $N=sn$; 
\item $s$ is the {\em sparsification factor} and is integer valued;
\item $f$ is the density of $\mathbf{\tilde{w}}^N$ (see~\eqref{eq:density});
\item $\mathbf{k}^N$ is a pseudo-random binary key sequence;
\item $\mathbf{x}^N$ is a binary sequence, generated  from the watermark $\mathbf{w}^n$ and the key $\mathbf{k}^N$, and embedded into flow IPDs;
\item $\mathbf{y}^{N'}$ is decoder's estimate of $\mathbf{x}^N$;
\item $\mathbf{\hat{w}}^n$ the estimate of the watermark sequence $\mathbf{w}^n$ at the decoder;
\item $\Delta$ is a real-value step size used for IPD quantizations. It represents the strength of the watermark signal;
\item $\sigma$ is the standard deviation of jitter;
\item $P_s, P_I,$ and $P_d$ represent the probability of a substitution, an insertion, and a deletion event in the communication channel model of the network, respectively. \end{itemize}

\section{The Proposed Scheme}
\label{sec:scheme}

    \begin{figure}[t]
       \centering
       \includegraphics[width=0.8\columnwidth]{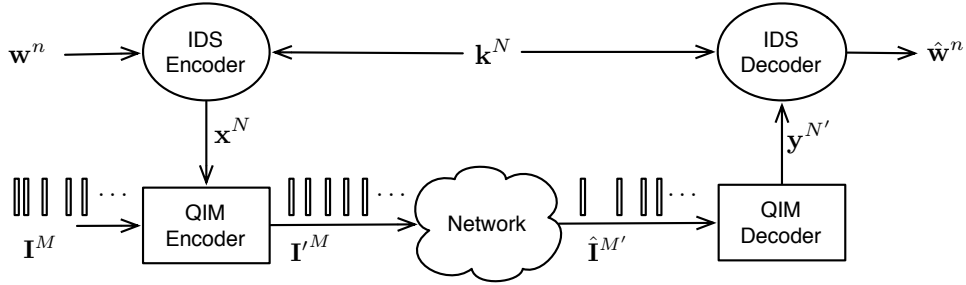}
       \caption{An overview of the proposed flow watermarking scheme. The watermark sequence $\mathbf{w}^n$ is first transformed into a codeword $\mathbf{x}^N$ with the help of the key $\mathbf{k}^{N}$.
$\mathbf{x}^N$ is then embedded into flow IPDs using QIM. 
At the decoder, the IPDs are processed by a QIM decoder to extract the codeword $\mathbf{y}^{N'}$, from which the IDS decoder recovers the watermark $\hat{\mathbf{w}}^N$, subsequently.}
       \label{fig:sys_mod}
    \end{figure}

 \begin{figure}[t] 
   \centering
   \includegraphics[width=0.8\columnwidth]{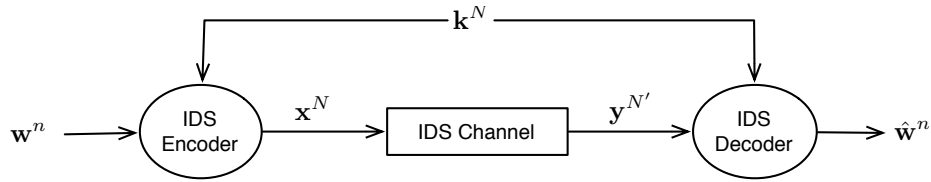}
   \caption{Abstraction of communication channel.
  The IDS encoder/decoder pair help correct the dependent substitution, deletion, and bursty insertion errors on the channel.}
   \label{fig:ids_channel}
\end{figure}

\subsection{Overview of the System}

Figure~\ref{fig:sys_mod} depicts the schematic of our proposed scheme, which can be divided into two layers: the insertion deletion substitution (IDS) encoder/decoder and the quantization index modulation (QIM) encoder/decoder.
In the upper layer,  the watermark sequence $\mathbf{w}^n$ is processed to generate an IDS error-correction codeword $\mathbf{x}^N$. 
On the lower layer, a QIM framework is used to inject  $\mathbf{x}^N$ into the IPDs of the flow.
QIM embedding is blind and causes little change to packet timings~\cite{Chen01}.
Upon receiving the flow, the QIM decoder  extracts the pattern $\mathbf{y}^{N'}.$ Subsequently an IDS decoder recovers the watermark, $\hat{\mathbf{w}}^n$, from this pattern. 

If we abstract the QIM encoder, the network, and the QIM decoder together as a channel, which takes $\mathbf{x}^N$ as the input and spits out $\mathbf{y}^{N'}$, 
flow watermarking is equivalent to solving the problem of sending one bit of information (the presence of the watermark) over this compound communication channel (see Figure~\ref{fig:ids_channel}). 
Codes for this compound channel  must withstand {\em dependent substitution, deletion, and bursty insertion} errors. We next introduce each component of our scheme in details.

\subsection{Insertion Deletion Substitution (IDS) Encoder}
\label{sec:ids_enc}

Our IDS error correction scheme is inspired by~\cite{Davey00, Davey01}, where a `marker' code is employed to provide reliable communications over a channel with deletion and insertion errors.
However, the approach in~\cite{Davey00, Davey01} is not directly applicable to our channel, 
as we need to deal with somewhat more complicated errors, such as dependent substitution, deletion, and bursty insertions which we discuss in~\S\ref{sec:error_analysis}.

The IDS encoder works as follows. 
The watermark sequence $\mathbf{w}^n$ is first sparsified into a longer sequence $\mathbf{\tilde{w}}^N$ of length $N=sn$, as given by
\begin{equation}
\tilde{w}_{(j-1)s+1}^{js} = S\left(w_j\right), \quad j=1,2,\cdots, n,
\label{eq:sparse}
\end{equation}
where $S(\cdot)$ is a deterministic sparsification function that pads $w_j$ with zeros, and is known at the decoder. 
We denote by density $f$ the ratio of `1' in $\mathbf{\tilde{w}}^N$, i.e., 
\begin{equation}
f =  \frac{\sum_{i=1}^{N}\tilde{w}_{i}}{N}.
\label{eq:density}
\end{equation}
$f$ is a decoding parameter shared with the IDS decoder. 
The sparsified watermark $\mathbf{\tilde{w}}^N$ is then added to a key $\mathbf{k}^N$ to form the codeword $\mathbf{x}^N$:
\begin{equation}
x_{i}=\tilde{w}_{i} \oplus k_{i}, \quad i=1,2,\cdots,N,
\label{eq:add}
\end{equation}
where $\mathbf{k}^N$ is pseudo-random {\em key} sequence which is also known at the decoder.

Let us work on a small example of embedding one bit of watermark $w_1=1$ in a length 8 sequence. First $w_1$ is sparsified into an 8 bit sequence $\mathbf{\tilde{w}}^8=10000000$ (the sparsification factor $s = 8$). Then we add this sparse sequence to the first 8 bit of our key, $\mathbf{k}^8=11111011$. The resulting codeword is $\mathbf{x}^8=01111011$. 
Because $\mathbf{x}^8$ is only different from the key at one position, the decoder could infer the positions of deleted or inserted bits by comparing the received codeword with the key. For instance, if the decoder receives a codeword $\mathbf{y}^7=0111011$, one bit shorter than the key, then it knows that most likely a bit `1' from the second run was lost during transmission. 
Based on this observation, a probabilistic decoder can be developed to fully recover embedded bits, as will be discussed in~\S\ref{sec:dec}. 

Since $\mathbf{\hat{w}}^N$ is sparse, the codeword $\mathbf{x}^N$ is close to the key, which is known at the IDS decoder.
 Therefore, the IDS encoding helps  synchronize the lost/inserted bits at the cost of information capacity over the channel, which  is not a concern for flow watermarking (see~\S\ref{sec:design}).


%
%

\subsection{Insertion Deletion Substitution (IDS) Channel}
\label{sec:qim_enc_dec}

\subsubsection{QIM Embedding}
The codeword $\mathbf{x}^N$ is injected into IPDs of the original flow using QIM embedding.
Given a quantization step size $\Delta$, the QIM encoder changes the IPD, $I_i$, into an even (or odd) multiplier of $\frac{\Delta}{2}$ given the embedded bit $x_i$ is a bit 0 (or 1).
The IPDs after modifications are given by
\begin{equation}
I'_i = \left\{
  \begin{array}{l l}
    \left \lceil \frac{\max\left(\sum_{j=1}^{i}I_j-\sum_{j=1}^{i-1}I'_j,0\right)}{\Delta} \right \rceil\Delta & \text{if } x_{i}=0,  \\
    \left(\left \lceil \frac{\max\left(\sum_{j=1}^{i}I_j-\sum_{j=1}^{i-1}I'_j,0\right)}{\Delta} \right \rceil+0.5\right)\Delta  &\text{if } x_{i}=1,\\
  \end{array}\right.
\label{eq:qim_enc}
\end{equation}
for $i=1,2,\cdots N$,
where the ceiling function describes the operation that adds minimum delays to $Packet$~$i$ to form the desired multiplier of $\frac{\Delta}{2}$. 

At the QIM decoder, each embedded bit is extracted based on whether a received IPD is closer to an even or odd quantizer, i.e., 
 \begin{equation}
y_i = \left\{
  \begin{array}{l l}
  \lfloor{\frac{2\hat{I}_{i}}{\Delta}}\rfloor \mod 2  \quad \text{if } \frac{2\hat{I}_{i}}{\Delta}-\lfloor\frac{2\hat{I}_{i}}{\Delta}\rfloor \le 0.5,  \\
    \lceil{\frac{2\hat{I}_{i}}{\Delta}}\rceil \mod 2 \quad \text{if } \frac{2\hat{I}_{i}}{\Delta}-\lfloor\frac{2\hat{I}_{i}}{\Delta}\rfloor > 0.5.\\
  \end{array}\right .
  \label{eq:qim_dec}
\end{equation}

\subsubsection{Channel Model}
\label{sec:error_analysis}
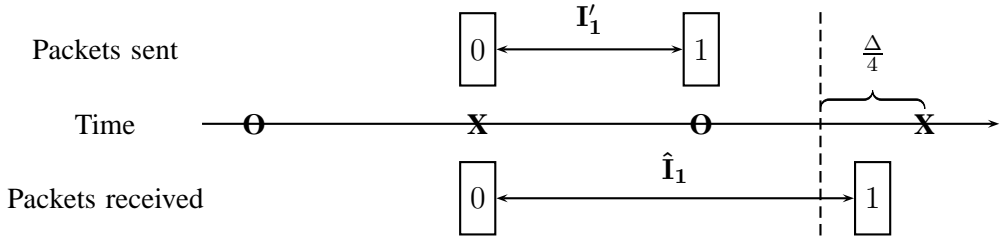
\begin{figure}[t]
\centering
\psset{unit=0.06\columnwidth}
\begin{pspicture}[showgrid=false](-2,-2)(13,2)

\newcommand{\prch}{-1.5} 
\newcommand{\prc}{-1} 
\newcommand{\prs}{1} 
\newcommand{\prsh}{1.5} 
\newcommand{\cl}{0} 
\newcommand{\clt}{0.2}

\newcommand{\co}{2}
\newcommand{\ct}{8}
\newcommand{\xo}{5}
\newcommand{\xt}{11}

\newcommand{\offset}{9.6} 
\newcommand{\npos}{10.3} 

\rput(0,\cl){\rnode{pt}\large{Time}}
\rput(0,\prc){\rnode{pt}\large{Packets received}}
\rput(0,\prs){\rnode{pt}\large{Packets sent}}

\dotnode[dotstyle=square*,dotscale=0.001](1.3,0){t1}
\dotnode[dotstyle=square*,dotscale=0.001](12,0){t2}
\ncline{->}{t1}{t2} \naput[npos=.5]{}

\rput(\co,\cl){\rnode{pt}\large{\textbf{O}}}
\rput(\xo,\cl){\rnode{pt}\large{\textbf{X}}}
\rput(\ct,\cl){\rnode{pt}\large{\textbf{O}}}
\rput(\xt,\cl){\rnode{pt}\large{\textbf{X}}}

\psfblock[framesize=0.5 1](\xo,\prs){rqp1}{\large{$0$}}
\psfblock[framesize=0.5 1](\ct,\prs){rqp2}{\large{$1$}}
\ncline{<->}{rqp1}{rqp2} \naput[npos=.5]{$\mathbf{I'_1}$}

\psfblock[framesize=0.5 1](\xo,\prc){rt1}{\large{$0$}}
\psfblock[framesize=0.5 1](\npos,\prc){rt2}{\large{$1$}}
\ncline{<->}{rt1}{rt2} \naput[npos=.5]{$\mathbf{\hat{I}_1}$}

\dotnode[dotstyle=square*,dotscale=0.001](\offset,\prsh){of1}
\dotnode[dotstyle=square*,dotscale=0.001](\offset,\prch){of2}
\ncline[style=Dash]{of1}{of2} \naput[npos=.5]{}

\psBraceUp[nodesepB=-.5](\xt,\clt)(\offset,\clt){$\frac{\Delta}{4}$}

\end{pspicture}
\caption{An example of substitution errors caused by network jitter. `$\mathbf{x}$'s denote even quantizers  and  `$\mathbf{o}$'s  odd quantizers. The bit embedded in $I_1$ is `1', but
the decoded bit from $\hat{I}_1$ (delay between received packets~$0$ and $1$) is `0'.}
\label{fig:jitter}
\end{figure}
In presence of network artifacts, received IPDs, $\hat{\mathbf{I}}^{M'}$, are different from the original IPDs ${\mathbf{I}}^M$, leading to errors in decoding $\mathbf{x}^N$. 
{\em Substitution} errors occur when network jitter alters IPDs largely. 
Figure~\ref{fig:jitter} depicts one example where an embedded bit is flipped by jitter. 
In Figure~\ref{fig:jitter}, the bit `$x_1=1$' was originally encoded in the IPD $I_1$, resulting in $I'_1=\frac{\Delta}{2}$.
But at the QIM decoder, the received IPD $\hat{I}_1$ is pushed by jitter into interval $(\frac{3\Delta}{4},\Delta)$, and thus decoded as `$y_1=0$'.
In absence of packet drops or splits, a watermark bit flips if the IPD jitter is  larger than $\frac{\Delta}{4}$.
Following the observation of previous work that shows IPD jitter  (within a certain period of time) is approximately  i.i.d. zero-mean Laplace distributed~\cite{Houmansadr09}, the probability of a substitution error by jitter can be estimated as
\begin{equation}
P_s= 1-F\left(\frac{\Delta}{4}\right) = \frac{1}{2}e^\frac{ {-\Delta} }{2\sqrt{2}\sigma},
\label{eq:substitution}
 \end{equation}
where $F(\cdot)$ is the Laplacian pdf and $\sigma^2$ is its variance.

\begin{figure}[t]
\centering
\psset{unit=0.06\columnwidth}
\begin{pspicture}[showgrid=false](-2,-3)(15,2)

\newcommand{\prc}{-1}
\newcommand{\prs}{1}
\newcommand{\pkz}{2}
\newcommand{\pko}{4}
\newcommand{\pkt}{7}
\newcommand{\pkth}{9.5}
\newcommand{\pkf}{11}
\newcommand{\pkfi}{13}
\newcommand{\spos}{0.3}
\newcommand{\apos}{-1.7}

\rput(0,0){\rnode{pt}\large{Time}}
\rput(0,\prc){\rnode{pt}\large{Packets received}}
\rput(0,\prs){\rnode{pt}\large{Packets sent}}

\dotnode[dotstyle=square*,dotscale=0.001](1.3,0){t1}
\dotnode[dotstyle=square*,dotscale=0.001](14,0){t2}
\ncline{->}{t1}{t2} \naput[npos=.5]{}

\dotnode[dotstyle=square*,dotscale=0.001](\pkz,\spos){p0}
\dotnode[dotstyle=square*,dotscale=0.001](\pko,\spos){p1}
\dotnode[dotstyle=square*,dotscale=0.001](\pkt,\spos){p2}
\dotnode[dotstyle=square*,dotscale=0.001](\pkth,\spos){p3}
\dotnode[dotstyle=square*,dotscale=0.001](\pkf,\spos){p4}
\dotnode[dotstyle=square*,dotscale=0.001](\pkfi,\spos){p5}

\ncline[linecolor=white]{p0}{p1} \ncput[npos=.5]{$x_1$}
\ncline[linecolor=white]{p1}{p2} \ncput[npos=.5]{$x_2$}
\ncline[linecolor=white]{p2}{p3} \ncput[npos=.5]{$x_3$}
\ncline[linecolor=white]{p3}{p4} \ncput[npos=.5]{$x_4$}
\ncline[linecolor=white]{p4}{p5} \ncput[npos=.5]{$x_5$}


\dotnode[dotstyle=square*,dotscale=0.001](\pkz,\apos){a0}
\dotnode[dotstyle=square*,dotscale=0.001](\pko,\apos){a1}
\dotnode[dotstyle=square*,dotscale=0.001](\pkt,\apos){a2}
\dotnode[dotstyle=square*,dotscale=0.001](\pkth,\apos){a3}
\dotnode[dotstyle=square*,dotscale=0.001](\pkf,\apos){a4}
\dotnode[dotstyle=square*,dotscale=0.001](\pkfi,\apos){a5}


\psfblock[framesize=0.5 1](\pkz,\prs){st1}{\large{$0$}}
\psfblock[framesize=0.5 1](\pko,\prs){st2}{\large{$1$}}
\psfblock[framesize=0.5 1](\pkt,\prs){st3}{\large{$2$}}
\psfblock[framesize=0.5 1](\pkth,\prs){st4}{\large{$3$}}
\psfblock[framesize=0.5 1](\pkf,\prs){st5}{\large{$4$}}
\psfblock[framesize=0.5 1](\pkfi,\prs){st6}{\large{$5$}}
\ncline{<->}{st1}{st2} \naput[npos=.5]{$\mathbf{I_1}$}
\ncline{<->}{st2}{st3} \naput[npos=.5]{$\mathbf{I_2}$}

\psfblock[framesize=0.5 1](\pkz,\prc){str1}{\large{$0$}}
\psfblock[style=Dash,framesize=0.5 1](\pko,\prc){str2}{\large{$1$}}
\psfblock[framesize=0.5 1](\pkt,\prc){str3}{\large{$2$}}
\psfblock[style=Dash,framesize=0.5 1](\pkth,\prc){str4}{\large{$3$}}
\psfblock[style=Dash,framesize=0.5 1](\pkf,\prc){str5}{\large{$4$}}
\psfblock[framesize=0.5 1](\pkfi,\prc){str6}{\large{$5$}}
\ncline{<->}{str1}{str3} \naput[npos=.5]{$\mathbf{I'_1}$}

\end{pspicture}
\caption{
Merging of IPDs as the result of packet drops. The deletion of Packet~1 merges the first two IPDs $I_1$ and $I_2$. and the deletions of Packet 3 and 4 merge $I_3$, $I_4$ and $I_5$.}
\label{fig:deletion}
\end{figure}
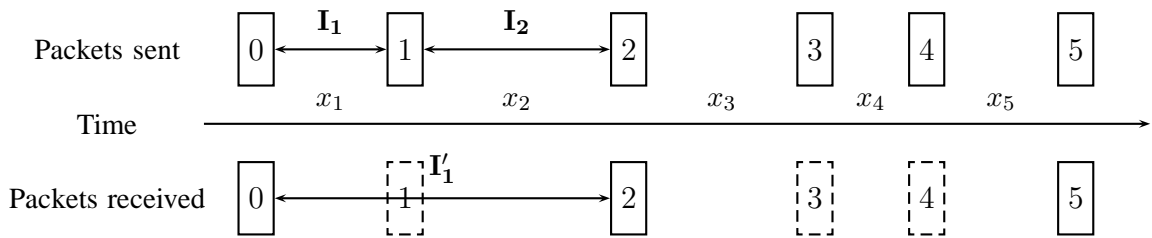

Decoding errors also occur when packets are dropped.
As packet drops lead to the merger of successive IPDs, the resulting error contains both deletion and substitution, which we refer to as {\em dependent deletion and substitution error}. 
For instance in  Figure~\ref{fig:deletion}, deletion of Packet~1 merges the IPDs $I_1$ and $I_2$ into a large received IPD $I'_1$.
As a result,  instead of $x_1$ and $x_2$, only one bit $x_1\oplus x_2$ is received at the decoder. 
We consider this case as a deletion of $x_1$, and possibly a substitution of $x_2$.
In this paper, we assume that each packet is dropped independently with probability  $P_d$.
For the convenience of analysis, we also assume that the head of watermarked packet sequence, $Packet$~$0$, is not dropped.

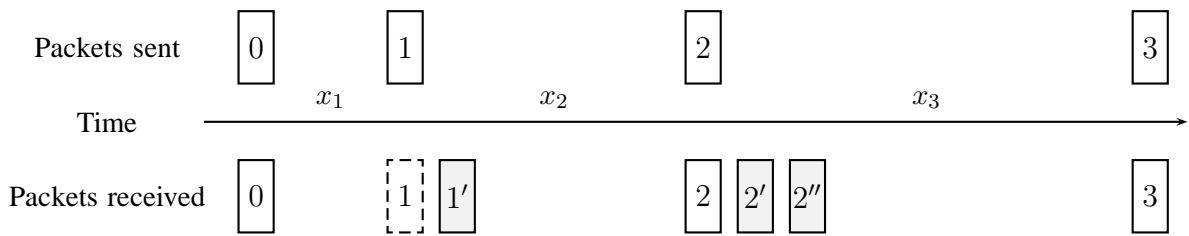
\begin{figure}[t]
\centering
\psset{unit=0.06\columnwidth}
\begin{pspicture}[showgrid=false](-2,-3)(15,2)

\newcommand{\prc}{-1}
\newcommand{\prs}{1}
\newcommand{\pkz}{2}
\newcommand{\pko}{4}
\newcommand{\pkt}{4.7}
\newcommand{\pkth}{8}
\newcommand{\pkf}{8.7}
\newcommand{\pkfi}{9.4}
\newcommand{\pks}{14}
\newcommand{\spos}{0.3}
\newcommand{\apos}{-1.7}

\rput(0,0){\rnode{pt}\large{Time}}
\rput(0,\prc){\rnode{pt}\large{Packets received}}
\rput(0,\prs){\rnode{pt}\large{Packets sent}}

\dotnode[dotstyle=square*,dotscale=0.001](1.3,0){t1}
\dotnode[dotstyle=square*,dotscale=0.001](14.5,0){t2}
\ncline{->}{t1}{t2} \naput[npos=.5]{}

\dotnode[dotstyle=square*,dotscale=0.001](\pkz,\spos){p0}
\dotnode[dotstyle=square*,dotscale=0.001](\pko,\spos){p1}
\dotnode[dotstyle=square*,dotscale=0.001](\pkt,\spos){p2}
\dotnode[dotstyle=square*,dotscale=0.001](\pkth,\spos){p3}
\dotnode[dotstyle=square*,dotscale=0.001](\pkf,\spos){p4}
\dotnode[dotstyle=square*,dotscale=0.001](\pkfi,\spos){p5}
\dotnode[dotstyle=square*,dotscale=0.001](\pks,\spos){p6}

\ncline[linecolor=white]{p0}{p1} \ncput[npos=.5]{$x_1$}
\ncline[linecolor=white]{p1}{p3} \ncput[npos=.5]{$x_2$}
\ncline[linecolor=white]{p3}{p6} \ncput[npos=.5]{$x_3$}

\dotnode[dotstyle=square*,dotscale=0.001](\pkz,\apos){a0}
\dotnode[dotstyle=square*,dotscale=0.001](\pko,\apos){a1}
\dotnode[dotstyle=square*,dotscale=0.001](\pkt,\apos){a2}
\dotnode[dotstyle=square*,dotscale=0.001](\pkth,\apos){a3}
\dotnode[dotstyle=square*,dotscale=0.001](\pkf,\apos){a4}
\dotnode[dotstyle=square*,dotscale=0.001](\pkfi,\apos){a5}
\dotnode[dotstyle=square*,dotscale=0.001](\pks,\apos){a6}


\psfblock[framesize=0.5 1](\pkz,\prs){st1}{\large{$0$}}
\psfblock[framesize=0.5 1](\pko,\prs){st1}{\large{$1$}}
\psfblock[framesize=0.5 1](\pkth,\prs){st1}{\large{$2$}}
\psfblock[framesize=0.5 1](\pks,\prs){st1}{\large{$3$}}

\psfblock[framesize=0.5 1](\pkz,\prc){str1}{\large{$0$}}
\psfblock[style=Dash,framesize=0.5 1](\pko,\prc){str1}{\large{$1$}}
\psfblock[FillColor=gray!10,framesize=0.5 1](\pkt,\prc){str1}{\large{$1'$}}
\psfblock[framesize=0.5 1](\pkth,\prc){str1}{\large{$2$}}
\psfblock[FillColor=gray!10,framesize=0.5 1](\pkf,\prc){str1}{\large{$2'$}}
\psfblock[FillColor=gray!10,framesize=0.5 1](\pkfi,\prc){str1}{\large{$2''$}}
\psfblock[framesize=0.5 1](\pks,\prc){str1}{\large{$3$}}

\end{pspicture}
\caption{A scenario with packet insertions. Packet 1 is split into two packets, and Packet~2 is split into three pieces.}
\label{fig:insertion}
\end{figure}

The last type of error comes from packet insertions. 
This happens when packets are split  to meet a smaller packet size limit, or when TCP transmission is triggered by network congestions.
Both cases cause bursty insertions of packets. 
An example of such a scenario is depicted in Figure~\ref{fig:insertion}. 
Packet 2 is split into three smaller ones, creating two new IPDs (2-2' and 2'-2'', both with zero length). Therefore,  two extra `0' bits would be decoded in $\mathbf{y}^{N'}$. 
In general, newly generated packets are mostly right next to the original one, hence we consider all inserted bits are`0's.\footnote{Our methodology can be extended to cover the case that both `0' and `1' bits may be inserted.}
Furthermore, we assume the number of inserted packets follows a geometric distribution with parameter $P_I$.

\subsection{Insertion Deletion Substitution (IDS) Decoder}
\label{sec:dec}
We estimate each watermark bit from $\mathbf{y}^{N'}$ using the maximum likelihood decoding rule given by 
\begin{equation}
\hat{w}_j = \arg \underset{w_j\in\{0,1\}}{\max} P\left(\mathbf{y}^{N'}|w_j\right), \quad j=1,2,\cdots, n.
\label{eq:ml}
\end{equation}

Since $\mathbf{x}^{N}$ is a deterministic function of $\mathbf{w}$, we derive the likelihood in~\eqref{eq:ml} based on the dependency between $\mathbf{y}^{N'}$ and $\mathbf{x}^{N}$ over the IDS channel. 
Suppose the QIM decoder received $i'-1$ packets after the first $i-1$ packets were sent out by the QIM encoder, and assume the $i'-1^{th}$ packet in the received flow corresponds to the $q^{th}$ packet in the sent flow or a packet inserted immediately after it ($q\leq i-1$).
The possible outcomes after $Packet$~$i$ is sent are:
\begin{itemize}
\item if $Packet$~$i$ in the sent flow is lost and no packets are inserted, the QIM decoder cannot decode new bits;
\item if $Packet$~$i$ is lost but $l>0$ new packets are inserted right after it, the decoder could decode $l$ bits, $y_{i'}^{i'+l-1}$, from newly received IPDs;
\item if $Packet$~$i$ is received and additionally $l\geq 0$ packets are inserted, the decoder can decode $l+1$ new bits, $y_{i'}^{i'+l}$.
\end{itemize}

In the last two cases, the new IPD between the $i'-1^{th}$ packet and $i'^{th}$ packets in the received flow corresponds to the merger of all IPDs between $Packet$~$q$ and $Packet$~$i$ in the sent flow. Hence, the first new bit, $y_{i'}$, is given by 
\begin{equation}
\begin{aligned}
y_{i'} =
&\begin{cases}
x_{q+1} \oplus x_{q+2} \cdots \oplus x_{i} & \text{w. p.}\quad1- P_s,\\
x_{q+1} \oplus x_{q+2} \cdots \oplus x_{i} \oplus 1 & \text{w. p.}\quad P_s,
\end{cases}
\end{aligned}
\label{eq:temp1}
\end{equation}
where $P_s$ is the probability of a substitution error given in~\eqref{eq:substitution}. 
The remaining new bits $y^{i'+l-1}_{i'+1}$ (or $y^{i'+l}_{i'+1}$) are just `0' bits resulting from bursty packet insertions. 

\subsubsection{Hidden Markov Model}

To capture the evolution of newly decoded bits from the received flow, we define the state after sending each packet with the pair $(x'_i,d_i)$, for $ i=1,2,\cdots N$, where
\begin{itemize}
\item The {\em accumulated bit} $x'_i$ it the sum of all bits resulting from merger of the IPDs between $Packet$~$i$ and the previous packet that was received at the decoder.
If $Packet$~$i-1$ was received, then the $x'_i$ is just  the bit embedded on the IPD between $Packet$~$i$ and~$i-1$, i.e., $x_i$. On the other hand,  if $Packet$~$i-1$ was completely lost (i.e., after its deletion, there were no insertions), $x'_i$ would be the sum of current bit $x_i$ and bits embedded on previously merged IPDs, i.e., $x_i\oplus x'_{i-1}$. To sum up, 
\begin{equation}
\begin{aligned}
x'_{i}=
&\begin{cases}
x_{i} & \text{w. p.} \quad 1-P_d\left(1-P_I\right),\\
x_{i}  \oplus x'_{i-1} & \text{w. p.} \quad P_d\left(1-P_I\right).\\
\end{cases}
\end{aligned}
\label{eq:accumulated_bit}
\end{equation} 
Recall from~\eqref{eq:add} that $\mathbf{x}^N$ is generatd using the key $\mathbf{k}^N$ and the sparse watermark sequence $\mathbf{\tilde{w}}$.
We will model the sparse watermark bits  $\tilde{w}_i$'s as independent $Bernoulli(f)$ random variables. Therefore \eqref{eq:accumulated_bit} can be rewritten as
\begin{equation}
\begin{aligned}
x'_{i}=
&\begin{cases}
k_{i} & \text{w. p.}  \quad (1-f)\left(1-P_d(1-P_I)\right),\\
k_{i} \oplus 1 & \text{w. p.} \quad f\left(1-P_d(1-P_I)\right),\\
 k_{i}  \oplus x'_{i-1} & \text{w. p.} \quad (1-f)P_d\left(1-P_I\right),\\
  k_{i}  \oplus x'_{i-1} \oplus 1 & \text{w. p.} \quad fP_d\left(1-P_I\right).
 \end{cases}
\end{aligned}
\label{eq:temp3}
\end{equation} 

Note that from~\eqref{eq:accumulated_bit}, we can rewrite~\eqref{eq:temp1} as
\begin{equation}
\begin{aligned}
y_{i'} =
&\begin{cases}
x'_{i}& \text{w. p.}\quad1- P_s,\\
x'_{i}\oplus1& \text{w. p.}\quad P_s.
\end{cases}\end{aligned}
\label{eq:temp1_1}
\end{equation}

\item The {\em drift} $d_i$ is the shift in position of the sent $Packet$~$i$ in the received flow, i.e., if   $Packet$~$i$ was not lost, it would appear at position $i'=i+d_i$ in the received flow.   
 Given $d_{i-1}$, the drift of $Packet$~$i$ is updated as 
 \begin{equation}
\begin{aligned}
d_{i}=
&\begin{cases}
d_{i-1}-1 & \text{w. p.}\quad P_{d}\left(1-P_I\right),\\
d_{i-1}+l, l\geq 0 & \text{w. p.}\quad \left(P_dP_I^{l+1}(1-P_I)+(1- P_d)P_I^{l}(1-P_I)\right),
\end{cases}
\end{aligned}
\label{eq:drift}
\end{equation}
where the first case occurs when $Packet$~$i-1$ was dropped with no new packets inserted, and the second case 
occurs when total of $l$ packets are received either because $Packet$~$i-1$ was dropped and there were $l+1$ insertions or  $Packet$~$i$ was received and there were $l$ insertions.  For the first Packet~0, we initialize $d_0=0$. This minor change let us relax the synchronization requirement on the first packet. 
\end{itemize}

Combine~\eqref{eq:temp1_1} and~\eqref{eq:temp3}, and given $i'=i+d_i$, we have 
\begin{equation}
\begin{aligned}
y_{i+d_{i}} =
&\begin{cases}
k_{i} & \text{w. p.} \quad \left((1-f)(1-P_s)+fP_s\right)\left(1-P_d(1-P_I)\right),\\
k_{i} \oplus 1 & \text{w. p.} \quad \left(f(1-P_s)+(1-f)P_s\right)\left(1-P_d(1-P_I)\right),\\
x'_{i-1}\oplus k_{i} & \text{w. p.} \quad \left((1-f)(1- P_s)+fP_s\right) P_d\left(1-P_I\right),\\
x'_{i-1} \oplus k_{i} \oplus 1 & \text{w. p.} \quad \left(f(1- P_s)+ (1-f)P_s\right) P_d\left(1-P_I\right).
\end{cases}
\end{aligned}
\label{eq:temp4}
\end{equation}

\begin{figure}[t]
\centering
\psset{unit=0.06\columnwidth}
\begin{pspicture}[showgrid=false](0,2)(15,6)

\newcommand{\cro}{0}
\newcommand{\crt}{4}
\newcommand{\crth}{8}
\newcommand{\crf}{12}
\newcommand{\rad}{.75}
\newcommand{\cy}{5}
\newcommand{\ccy}{4}

\newcommand{\crmido}{2}
\newcommand{\crmidt}{6}
\newcommand{\crmidth}{10}
\newcommand{\crmidf}{13.5}
\newcommand{\ly}{3}

\pszero[zeroradius=\rad](\cro,\cy){d1}
\pszero[zeroradius=\rad](\crt,\cy){d2}
\pszero[zeroradius=\rad](\crth,\cy){d3}
\pszero[zeroradius=\rad](\crf,\cy){d4}


\nclist{->}{ncline}{d1,d2,d3,d4}
\psset{arcangle=-43}
\nclist{->}{ncarc}{d1,d2,d3,d4}


\rput(d1){$x'_1,d_1$}
\rput(d2){$x'_2,d_2$}
\rput(d3){$x'_3,d_3$}
\rput(d4){$x'_4,d_4$}

\dotnode[dotstyle=square*,dotscale=0.001](\crmido,\ccy){p10}
\dotnode[dotstyle=square*,dotscale=0.001](\crmido,\ly){p11}

\dotnode[dotstyle=square*,dotscale=0.001](\crmidt,\ccy){p20}
\dotnode[dotstyle=square*,dotscale=0.001](\crmidt,\ly){p21}

\dotnode[dotstyle=square*,dotscale=0.001](\crmidth,\ccy){p30}
\dotnode[dotstyle=square*,dotscale=0.001](\crmidth,\ly){p31}

\dotnode[dotstyle=square*,dotscale=0.001](\crmidf,\cy){p40}
\dotnode[dotstyle=square*,dotscale=0.001](\crmidf,\ly){p41}
\ldotsnode(p40){fun2}
\ldotsnode(p41){fun2}

\dotnode[dotstyle=square*,dotscale=0.001](\cro,3.5){p00}

\ncline{->}{p10}{p11}
\ncline{->}{p20}{p21}
\ncline{->}{p30}{p31}

\ncline{->}{d1}{p00}

\rput(p00){$\{y_1,\dots,y_{d_1}\}$}
\rput(p11){$\{y_{d_1+1},\dots,y_{d_2+1}\}$}
\rput(p21){$\{y_{d_2+2},\dots,y_{d_3+2}\}$}
\rput(p31){$\{y_{d_3+3},\dots,y_{d_4+3}\}$}
\rput(p31){$\{y_{d_3+3},\dots,y_{d_4+3}\}$}

\end{pspicture}
\caption{The hidden-Markov model for the IDS channel. The observations are the codewords ($y_i$'s) received by the IDS decoder, and the hidden states keep track of the drift and accumulated bit when sending every packet.}
\label{fig:hmm}
\end{figure}
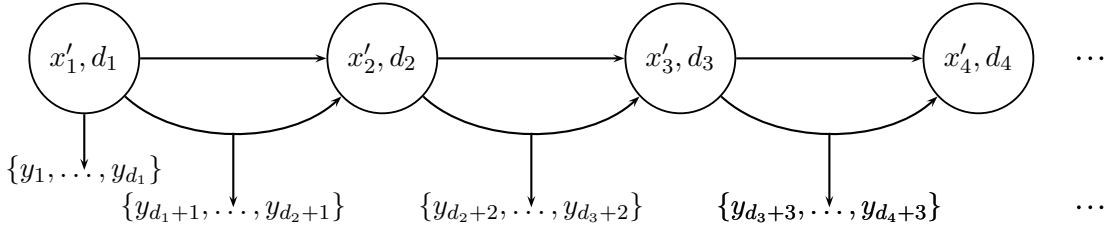

Equation~\eqref{eq:temp4} captures the HMM with  hidden states of $(x'_i,d_i), i=1,2,\cdots N$ and observation states of $\mathbf{y}^{N'}$, as depicted in Figure~\ref{fig:hmm}.
The state transition probabilities $P\left(y_{i-1+d_{i-1}}^{i-1+d_{i}}, x'_{i},d_i | x'_{i-1},d_{i-1}\right)$ can be derived using~\eqref{eq:temp3},~\eqref{eq:drift} and~\eqref{eq:temp4}, summarized as
\begin{equation}
\begin{aligned}
&P\left(y_{i-1+d_{i-1}}^{i-1+d_{i}}, x'_{i},d_i | x'_{i-1},d_{i-1}\right)=\\
&\begin{cases}
(1-f)P_d(1-P_I)& \text{if } x'_i=x'_{i-1}\oplus k_i, d_{i}=d_{i-1}-1 \text{ and } y_{i-1+d_{i-1}}^{i-1+d_i}=\emptyset,\\ 
fP_d(1-P_I)  & \text{if } x'_i=x'_{i-1}\oplus k_i\oplus1, d_{i}=d_{i-1}-1 \text{ and } y_{i-1+d_{i-1}}^{i-1+d_i}=\emptyset,\\
(1-f)(1-P_s)(1-P_I)( P_dP_I^{l+1}+(1-P_d)P_I^{l})& \text{if } x'_i=k_i, d_{i}=d_{i-1}+l \text{ and } y_{i-1+d_{i-1}}=x'_{i-1},\\
 f(1-P_s)(1-P_I)( P_dP_I^{l+1} +(1-P_d)P_I^{l})& \text{if } x'_i=k_i\oplus1, d_{i}=d_{i-1}+l \text{ and } y_{i-1+d_{i-1}}=x'_{i-1},\\
(1-f)P_s(1-P_I)( P_dP_I^{l+1}+(1-P_d)P_I^{l})& \text{if } x'_i=k_i, d_{i}=d_{i-1}+l \text{ and } y_{i-1+d_{i-1}}=x'_{i-1}\oplus1,\\
fP_s(1-P_I)( P_dP_I^{l+1}+(1-P_d)P_I^{l})& \text{if } x'_i=k_i\oplus1, d_{i}=d_{i-1}+l \text{ and } y_{i-1+d_{i-1}}=x'_{i-1}\oplus1.
\end{cases}
\end{aligned}
\label{eq:transition_prob}
\end{equation}

For example, after sending $Packet$~$i-1$, the system state is $(x'_{i-1},d_{i-1})$. If $Packet$~$i-1$ is lost and no packets are inserted. Then from~\eqref{eq:drift}, the drift of $Packet$~$i$ becomes  $d_{i}=d_{i-1}-1$, and no new bit is decoded, i.e., 
$y_{i-1+d_{i-1}}^{i-1+d_{i}}$ is an empty sequence.
Additionally, the IPD between $Packet$~$i$ and $i-1$ is added to previously merged IPDs such that $x'_{i}$ is decided based on the last two cases in~\eqref{eq:temp3}. 
Overall, the transition probability in this scenario is given by
\begin{equation}
\begin{aligned}
P\left(\emptyset, x'_{i}, d_{i-1}-1| x'_{i-1}, d_{i-1} \right)=
&\begin{cases}(1-f)P_d(1-P_I) & \text{if} \quad x'_{i}=x'_{i-1}\oplus k_{i},\\
fP_d(1-P_I) & \text{if} \quad x'_{i}=x'_{i-1}\oplus k_{i}\oplus 1.
\end{cases}
\end{aligned}
\end{equation}

\subsubsection{Forward-Backward Algorithm} For the HMM in Figure~\ref{fig:hmm}, we apply the forward-backward algorithm to derive the posterior probabilities $P(\mathbf{y}^{N'}|w_j)$, $j=1,2,\cdots, n$.
Let us define the \emph{forward} quantity as the joint probability of bits $y_1^{i-1+d_i}$ decoded before sending $Packet$~$i$ at the hidden state of $(x'_{i},d_i)$, which is given by
\begin{equation}
 \begin{aligned}
        F_{i}(x'_{i},d_i) = P(y_1^{i-1+d_i}, x'_{i},d_i), \quad i=1,2,\cdots, N.
       \end{aligned}
       \label{eq:forward_def}
\end{equation}
The forward quantities  can be computed  recursively using transition probabilities in~\eqref{eq:transition_prob} as  
\begin{equation}
 \begin{aligned}
      F_{i}(x'_{i},d_i)=\sum_{\substack{x'_{i-1},\\d_{i-1}}}F_{i-1}(x'_{i-1},d_{i-1}) P(y_{i-1+d_{i-1}}^{i-1+d_{i}}, x'_{i},d_i | x'_{i-1},d_{i-1}).
        \end{aligned}
       \label{eq:forward}
\end{equation}

Similarly,  we define the \emph{backward} quantity as the conditional probability of decoding the rest  of the bits in the received flow, $y_{i+d_i}^{N'}$, given the current state $(x'_{i},d_i)$,
\begin{equation}
 B_{i}(x'_{i},d_i)=P(y_{i+d_i}^{N'}|x'_{i},d_i), \quad i=1,2,\cdots, N.
 \label{eq:backward_def}
\end{equation}

The backward quantities can also be computed recursively as 
\begin{equation}
\begin{aligned}
B_{i}(x'_{i},d_i)=\sum_{\substack{x'_{i+1},\\d_{i+1}}}P(y_{i+d_{i}}^{i+d_{i+1}}, x'_{i+1},d_{i+1} | x'_{i},d_{i})B_{i+1}(x'_{i+1},d_{i+1}).
       \end{aligned}
       \label{eq:backward}
\end{equation}

Given  the forward/backward quantities, the posterior likelihood of the watermark bit $w_j$ is  given by
\begin{equation}
 \begin{aligned} 
 P\left(\mathbf{y}^{N'}|w_j\right) &= P\left(\mathbf{y}^{N'}|\tilde{w}^{js}_{(j-1)s+1}\right)\\
 &= \sum_{\substack{x'_{(j-1)s},x'_{js},\\d_{(j-1)s},d_{js}}} F_{(j-1)s}\left(x'_{(j-1)s},d_{(j-1)s}\right)\hat{F}_{js}\left(x'_{js},d_{js})B_{js}(x'_{js},d_{js}\right),
       \end{aligned}
       \label{eq:post}
\end{equation}
where the first equality follows from our watermark sparsification function in~\eqref{eq:sparse}, and the quantity $F'_{js}(x_{i},d_{i})$  is defined as
\begin{equation}
\hat{F}_{js}(x'_{i},d_{i})=P\left(y_{(j-1)s+d_{(j-1)s}}^{i-1+d_{i}},x'_{i},d_{i}|x'_{(j-1)s},d_{(j-1)s},\tilde{w}^{js}_{(j-1)s+1}\right), \quad (j-1)s+1 \leq i\leq js.
\end{equation} 

The quantity $F'_{js}(x'_{i},d_{i})$ can be calculated recursively as 
\begin{equation}
 \begin{aligned}
 \hat{F}_{js}(x'_{i},d_{i})= \sum_{\substack{x'_{i-1},\\d_{i-1}}}\hat{F}_{js}(x'_{i-1},d_{i-1})P\left(y_{i-1+d_{i-1}}^{i-1+d_{i}},x'_{i},d_{i}|x'_{i-1},d_{i-1},\tilde{w}^{js}_{(j-1)s+1)}\right),
 \end{aligned}
  \label{eq:forward2}
\end{equation}
where  $P\left(y_{i-1+d_{i-1}}^{i-1+d_{i}},x'_{i},d_{i}|x'_{i-1},d_{i-1},\tilde{w}^{js}_{(j-1)s+1)}\right)$ is given by
\begin{equation}
\begin{aligned}
&P\left(y_{i-1+d_{i-1}}^{i-1+d_{i}},x'_{i},d_{i}|x'_{i-1},d_{i-1},\tilde{w}^{js}_{(j-1)s+1)}\right)=\\
&\begin{cases}
P_d(1-P_I) \quad \text{if }  d_{i}=d_{i-1}-1 \text{ and } x'_{i}=\tilde{w}_{i}\oplus k_{i}\oplus x'_{i-1},  \text{ and } y_{i-1+d_{i-1}}^{i-1+d_i}= \emptyset,\\
P_s (1-P_I)\left(P_dP_I^{d_i-d_{i-1}+1}+(1-P_d)P_I^{d_i-d_{i-1}}\right) \quad \text{if } d_{i}\geq d_{i-1}, x'_{i}=\tilde{w}_{i}\oplus k_{i} \text{ and } y_{i-1+d_{i-1}}= x'_{i-1}\oplus 1,\\
(1-P_s)(1-P_I) \left(P_dP_I^{d_i-d_{i-1}+1}+(1-P_d)P_I^{d_i-d_{i-1}}\right)
\quad \text{if } d_{i}\geq d_{i-1},  x'_{i}=\tilde{w}_{i}\oplus k_{i} \text{ and } y_{i-1+d_{i-1}}= x'_{i-1}.
\end{cases}
\end{aligned}
\label{eq:sum_2}
\end{equation}

Once the posterior probabilities for all watermark bits are calculated, the 
watermark sequence, $\hat{\mathbf{w}}^n$ can be estimated using maximum likelihood rule of~\eqref{eq:ml}.
Finally, the presence of the watermark in a flow is decided based on the correlation value of the estimated watermark, $\hat{\mathbf{w}}^n$, and the original watermark sequence, $\mathbf{w}^n$.

\begin{table}[t]
\centering  \caption{True positive rates with varying watermark parameters when false positive rate is fixed below 1\%.}
\begin{tabular}{|l|c|c|c|c|c|}
\hline
\backslashbox{\quad  n}{$\Delta$ (ms)}&20&60&100\\
\hline  
    10 &  0.0310& 0.6050    &0.6224     \\
    30 &   0.0310& 0.9790    &0.9970     \\
    50 &      0.0272& 0.9990    &  1  \\
    \hline
\end{tabular}
  \label{tab:partesttp}
\end{table}
\begin{table}[t]
\centering
  \caption{True positive rates under varying IPD jitter with false positive rates below 1\%.}
     \begin{tabular}{|l|l|c|c|c|c|c|}
      \hline
      {Jitter Std. Dev. (ms)}&10&20&30&40\\
      \hline
      Synthetic traffic &1.000&0.989&0.770&0.232\\
            Real traffic &1.000&0.989&0.652&0.193\\
      \hline
    \end{tabular}
  \label{tab:jitter}
\end{table}
\begin{table}[t]
\centering
  \caption{True positive rates for varying $P_{d}$ with false positive rates below 1\%.}
     \begin{tabular}{|l|l|c|c|c|c|}
      \hline
      {$P_d$}&1\%&2\%&3\%&10\%\\
      \hline
       Synthetic traffic &1.000&1.000&1.000&0.995\\
             Real traffic & 1.000&1.000&1.000&0.996\\\hline 
    \end{tabular}

  \label{tab:deletion}
\end{table}
\begin{table}[t]
\centering
  \caption{True positive  rates for varing $P_{I}$ with false positive rates below 1\%.}
\begin{tabular}{|l|l|c|c|c|c|}
      \hline
      {$P_{I}$}&1\%&5\%&10\%&20\%\\
      \hline
    Synthetic traffic &1.000&1.000&1.000&0.500\\
             Real traffic &1.000&1.000&1.000&0.568\\
      \hline
    \end{tabular}
  \label{tab:insertion}
\end{table}
\begin{table}[!htbp]
\caption{True positive  rates for varying $P_{I}$,$P_{d}$ with false positive rates below 1\%.}
\centering
\begin{tabular}{|l|l|c|c|c|}
      \hline
      {$P_{i}$,$P_{d}$}&1\%,1\%&5\%,5\%&10\%,10\%\\
      \hline
      Synthetic&1.000&1.000&0.764\\
            Real&1.000&1.000&0.662\\
              \hline
        \end{tabular}
    \label{tab:insertiondeletion}
\end{table}

 \section{Evaluation}
\label{sec:eval}

We tested our scheme for two groups of traces: {\em synthetic} packet flows of length 2000 generated from Poisson process with average rate of $3.3$~packets per second (pps), and real {\em SSH} traces of length 2000 collected in CAIDA database with average rate of $0.865$~pps~\cite{CAIDA}, which represent typical traffic in human-involved network connections, where flow watermarks are most applicable. 

\subsection{Parameter Selection} 

The first test examined the effects of watermark length $n$ and IPD quantization step size $\Delta$.
We varied $n$ over $\{10,30,50\}$,  $\Delta$ over $\{20, 60,100\}$\,ms and fixed the sparisificatoin factor $s=10$. The deletion and insertion probabilities and the network jitter were set to 
 $P_d=0.1$, $P_I=0$, $\sigma=10$\,ms, respectively.  
5000 synthetic flows were embedded with watermarks and another 5000 unwatermarked ones served as the control group.

Table~\ref{tab:partesttp} shows the true positive rates of our test, when false positive rates were kept under 1\%. 
As we increase watermark length or quantization step size (embed a `stronger' pattern),  detection error decreases. 

{For the tests in this section, we fix the watermark parameters to $\{\Delta=100\,ms, n=50,  s=10\}$, which had the best performance in Table~\ref{tab:partesttp}}.

\begin{table}[t]
  \centering
    \caption{Average KS distances between watermarked and unwatermarked synthetic traces.}
    \begin{tabular}{|c|c|c|c|}
    \hline
            \backslashbox{\quad  n}{$\Delta$ (ms)}
           &100&80&60\\
            \hline
            30 &0.0177    &0.0138    &0.0101   \\
            \hline
            40  &0.0233    &0.0181    &0.0133   \\
            \hline
            50  &0.0284    &0.0223    &0.0160   \\
    \hline
    \end{tabular}
  \label{tab:KStest}  
\end{table}
\begin{table}[!htbp]
  \centering
    \caption{Average KS distances between watermarked and unwatermarked SSH traces.}
    \begin{tabular}{|c|c|c|c|}
    \hline
            \backslashbox{\quad  n}{$\Delta$ (ms)}
           &100&80&60\\
            \hline
            30 &0.0091    &0.0081    &0.0071   \\
            \hline
            40  &0.0120    &0.0111    &0.0091   \\
            \hline
            50  &0.0158    &0.0139    &0.0123   \\
    \hline
    \end{tabular}
  \label{tab:KStestssh} 
\end{table}

\subsection{Robustness Tests}

We evaluated watermark robustness against network jitter, and packet loss and insertion.

\subsubsection{IPD jitter}
\label{sec:jitter_tests}

From the experimental results in~\cite{Swirl},  the standard deviation of the Laplacian jitter is estimated as $\sigma=10$\,ms.  
We performed tests with $\sigma$ varied over $\{10, 20, 30, 40\}$\,ms.
The packet drop and split probabilities were $P_d=0.1$ and $P_I=0$, respectively.
This time, we watermarked 1000 flows from both synthetic and SSH traces. 

The true positive rates are given in Table~\ref{tab:jitter}.
Notice that the watermarks were detected with accuracies over $98\%$, even when jitter was as high as $20$\,ms. 
The detection performance falls sharply when jitter standard deviations exceeds $40$\,ms. 
However, such excessively large jitter rarely occurs at proper network conditions. Hence, our scheme  withstands network jitter in normal operating conditions.

\subsubsection{Packet deletion and insertion}
One major improvement of our design over previous work is robustness against packet deletion and insertion. 
To verify this, we tested our scheme in a network with: solely packet deletion with probabilities  $P_d=\{0.01,0.02,0.03,0.1\}$, solely packet insertion with probabilities $P_I=\{0.01,0.05,0.1,0.2\}$, and both deletion and  insertion with probabilities $(P_d,P_I)=\{(0.01,0.01),(0.05,0.05),(0.1,0.1),(0.15,0.15)\}$.
During all the tests, the standard deviation of jitter was fixed as $\sigma =10\,ms$, and 1000 flows from both synthetic and SSH traces were used. 

The results  in Tables~\ref{tab:deletion}--\ref{tab:insertiondeletion} demonstrate  watermarks were detected with high accuracies when  5\% of packets were dropped and inserted. 
\begin{table}[t]
  \centering
    \caption{Statistics of blank intervals in the aggregated flow from synthetic traces.}
    \begin{tabular}{|c|c|c|}
      \hline
       &Watermarked&Unwatermarked\\
      \hline
       Mean&24.07&25.96\\
      \hline
       Standard Deviation&5.246&5.187\\
      \hline
    \end{tabular}
  \label{tab:MultiFlow}
\end{table}
\begin{table}[t]
  \centering
    \caption{Statistics of blank intervals in the aggregated flow from SSH traces.}
    \begin{tabular}{|c|c|c|}
      \hline
       &Watermarked&Unwatermarked\\
      \hline
       Mean&403&395.67\\
      \hline
       Standard Deviation&20.54&16.84\\
      \hline
    \end{tabular}
  \label{tab:MFAssh}
\end{table}
\begin{figure}[t]
   \centering
   \subfigure[Unwatermarked flows]{
   \includegraphics[width=0.48\columnwidth]{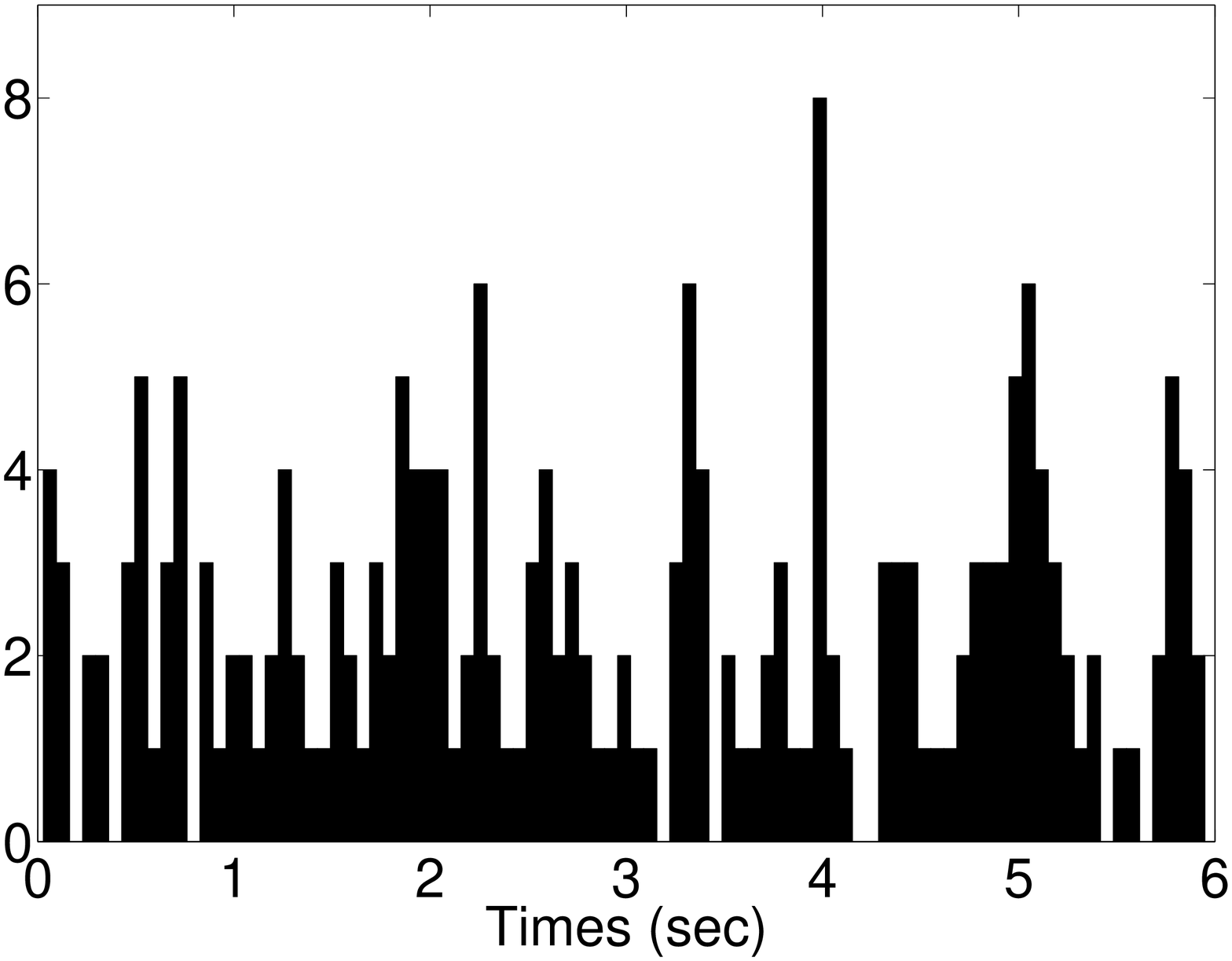}
   \label{fig:unmarked}
   }\subfigure[Watermarked flows]{
   \includegraphics[width=0.48\columnwidth]{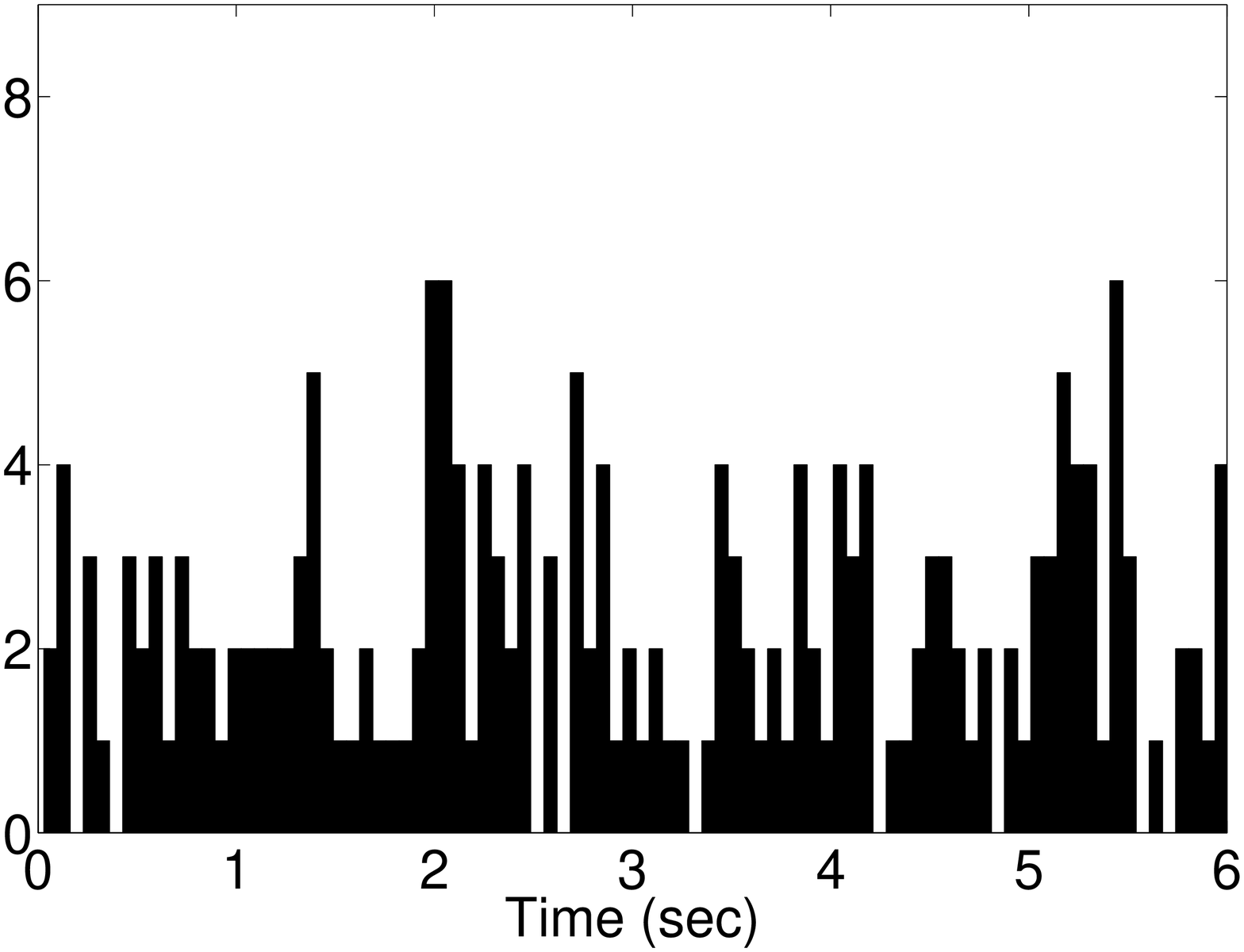}
   \label{fig:marked}
   }
   \caption{MFA tests on synthetic traces: packet counts in intervals of 70\,ms in the flow aggregated from (a) 10 unwatermarked flows and (b) 10 watermarked flows.}
   \label{fig:MultiFlow}
\end{figure}

\begin{figure}[t]
   \centering
   \subfigure[Unwatermarked flows]{
   \includegraphics[width=0.48\columnwidth]{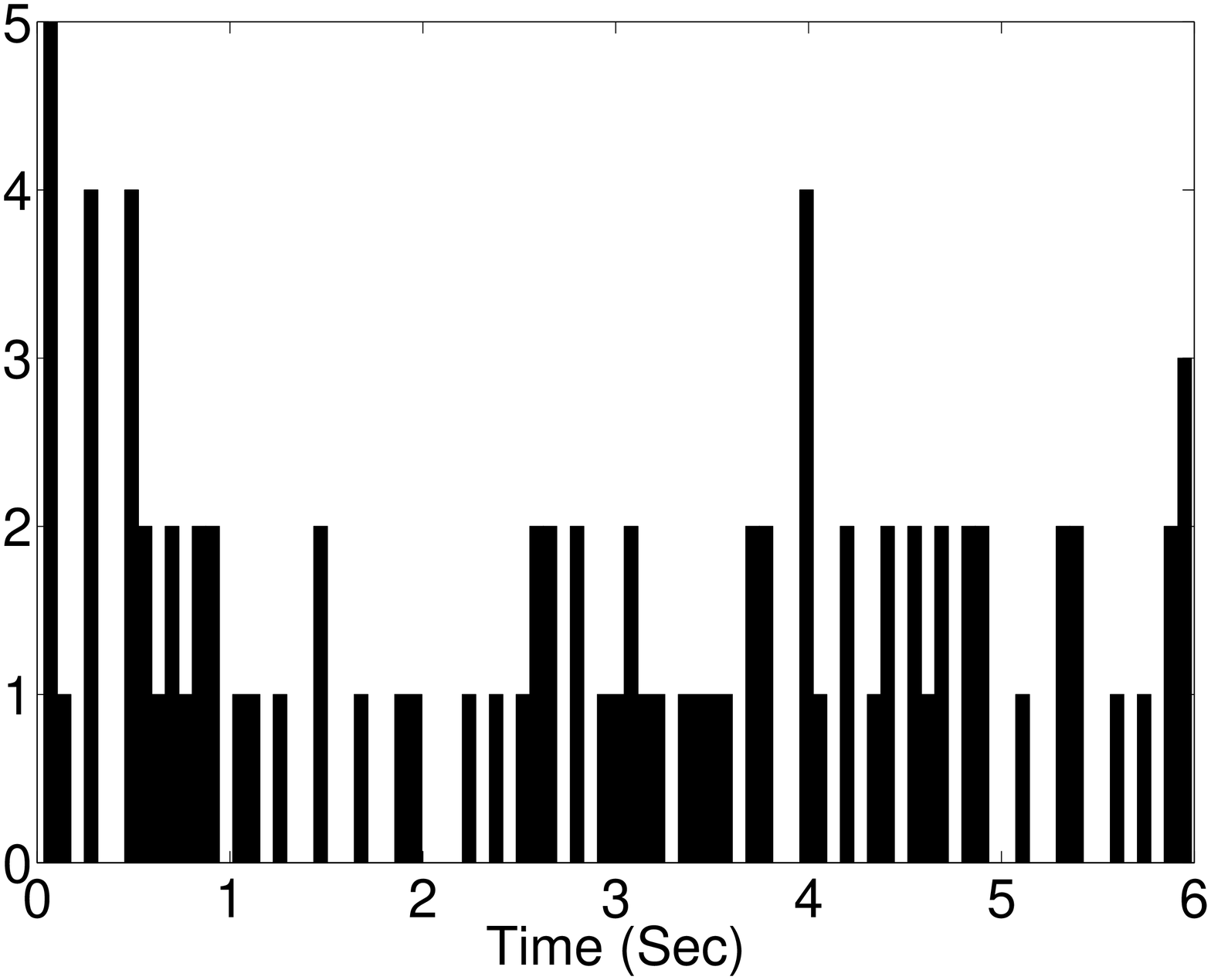}
   \label{fig:unmarkedssh}
   }\subfigure[Watermarked flows]{
   \includegraphics[width=0.48\columnwidth]{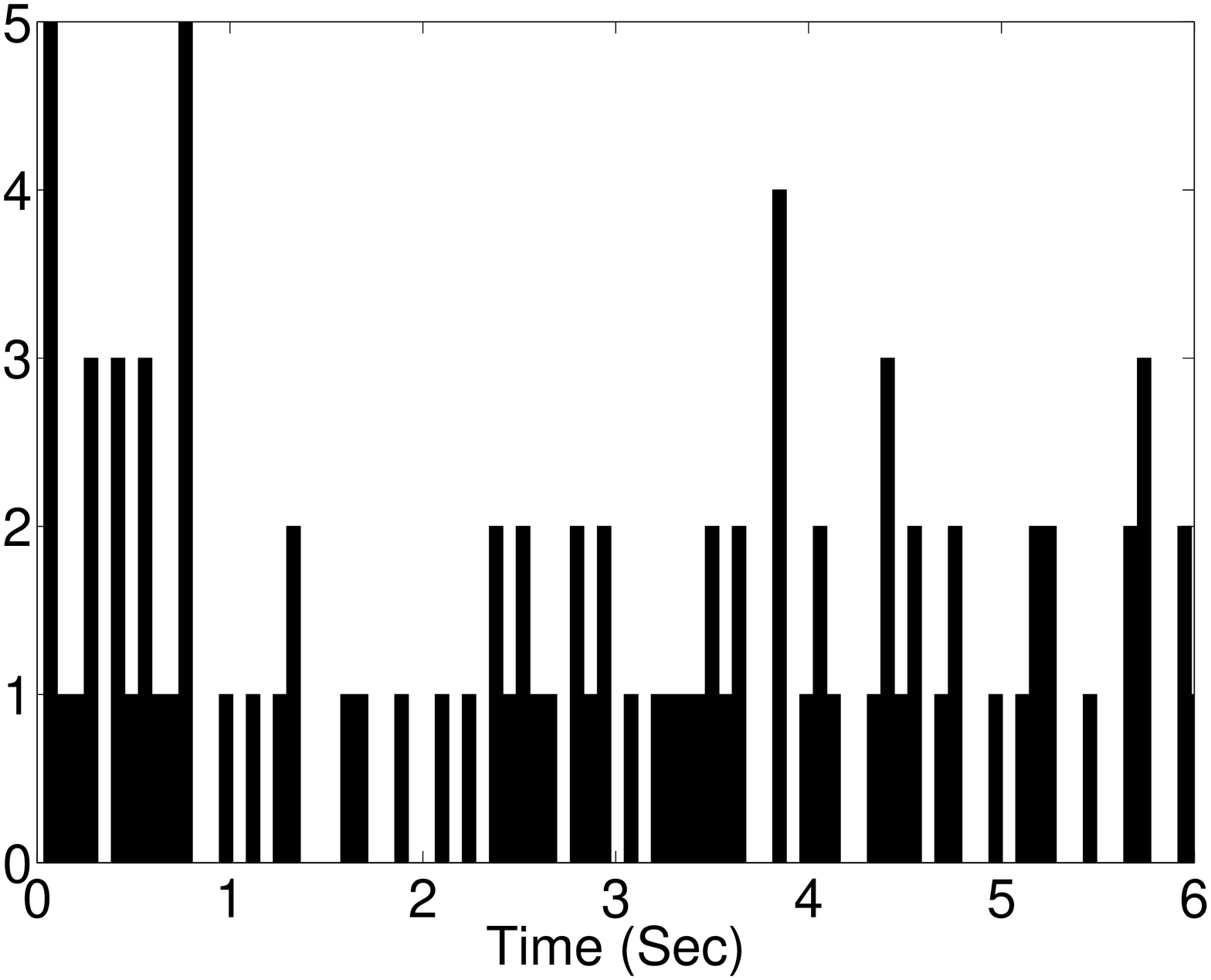}
   \label{fig:markedssh}
   }
   \caption{MFA tests on real SSH traces: packet counts in intervals of 70\,ms in the flow aggregated from (a) 10 unwatermarked flows and (b) 10 watermarked flows.}   \label{fig:MFAssh}
\end{figure}

\subsection{Visibility Tests}
\label{sec:visibility}
We first evaluated watermark invisibility with two Level-I attack tests: 
the {Kolmogorov-Simirnov (KS) test} and the  {multiflow attack (MFA) test}.

KS test is commonly applied to comparing distributions of datasets.
Given two data sets, the KS distance is computed as the maximum difference of their empirical distribution functions~\cite{Massey51}. 
For two flows $A$ and $B$, the KS distance is given by 
$\underset{x}{\sup}(|F_{A}(x)-F_{B}(x)|)$, 
where $F_A(x)$ and $F_B(x)$ are the empirical pdfs of IPDs in $A$ and $B$.
We claim two flows are indistinguishable if their KS distance is below 0.036, a threshold suggested in~\cite{Massey51}.
We calculated the average $KS$ distance between watermarked and unwatermarked flows using both synthetic and SSH traces. 
The results are tabulated in  Tables~\ref{tab:KStest} and~\ref{tab:KStestssh}.
None of the KS distances exceed the detection threshold of visibility, which implies the embedded watermarks did not cause noticeable artifacts in the original packet flows.

MFA is a watermark attack that detects positions of embedded watermarks in interval-based schemes, 
When flows which were watermarked using the same watermark are aggregated, the aggregate flow shows a number of 
  intervals containing no packets (see Figure~10 in~\cite{Kiyavash08}). 
To test whether such `visible' pattern exists in flows watermarked using our scheme, we combined 10 watermarked and 10 unwatermarked flows for both the synthetic and SSH traces, and divided the aggregated flows into intervals with length of  length of 70\,ms.
We then counted the number of blank intervals with no packets in each aggregate flow. 
This procedure was repeated 1000 times, and the resulting blank interval statistics are shown in Tables~\ref{tab:MultiFlow} and~\ref{tab:MFAssh}. For both synthetic and SSH traces, we see that the number of blank intervals does not change much after watermarks were embedded. 
Figure~\ref{fig:MultiFlow} depicts packet counts in each interval of length 70~ms in the aggregated synthetic traces. 
Comparing Figures~\ref{fig:unmarked} with~\ref{fig:marked}, no clear watermark pattern is observed.
The same observation was made in  Figure~\ref{fig:MFAssh}, which depicts packet counts of SSH traces. 
Therefore, our scheme is resistant to MFA. 

We next tested the performance of our watermarks under a Level-II attack, BACKLIT, where the attacker sees both directions of a TCP connection~\cite{Luo}.
BACKLIT detects watermarks in SSH flows based on the differences in round trip times (RTTs) of consecutive TCP requests, $\Delta RTT$.
We considered a stepping stone detection scenario in our campus network.
Network jitter in such an environment, like most enterprise networks, is very small.
According to our measurements from our lab machine to the campus exit node, the jitter standard deviation was as low as $\delta=1.6$\,ms. 
For this level of noise, a small quantization step size of IPDs, 10\,ms, was sufficient to achieve accurate decoding performances (true positive rate of $100$\% and false positive rate of  less than 1\%).
We then examined the effect of our watermarks on the $\Delta RTT$ distribution of a SSH connection, by monitoring the RTT jitter from our lab to 5 PlanetLab nodes~\cite{Planetlab}.
For each node, we issued  4000 ping requests with ping interval of 100\,ms, and 
divided ping packets  into two windows, each consisting of 2000 packets. 
We transplanted delays of SSH packets during watermarking onto the ping replies in Window-1, to mimic the effects of watermarking live TCP requests, and left Window-2 untouched as the control group. 
\begin{figure}[t]
   \centering
   \subfigure[PlanetLab node: 13.7.64.22]{
   \includegraphics[width=0.32\columnwidth]{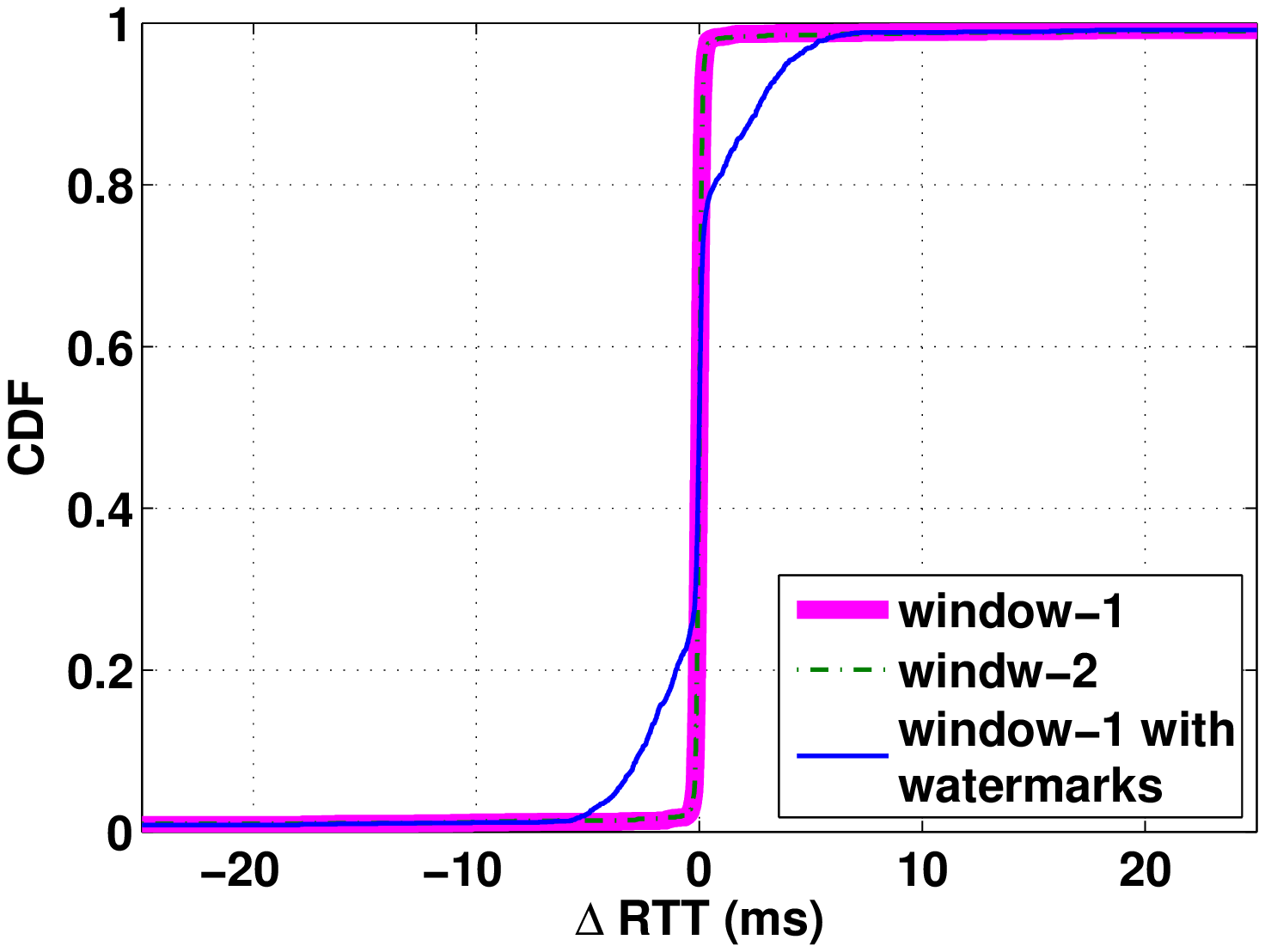}
   \label{fig:host1}
   }\subfigure[PlanetLab node: 130.65.6.225]{
   \includegraphics[width=0.32\columnwidth]{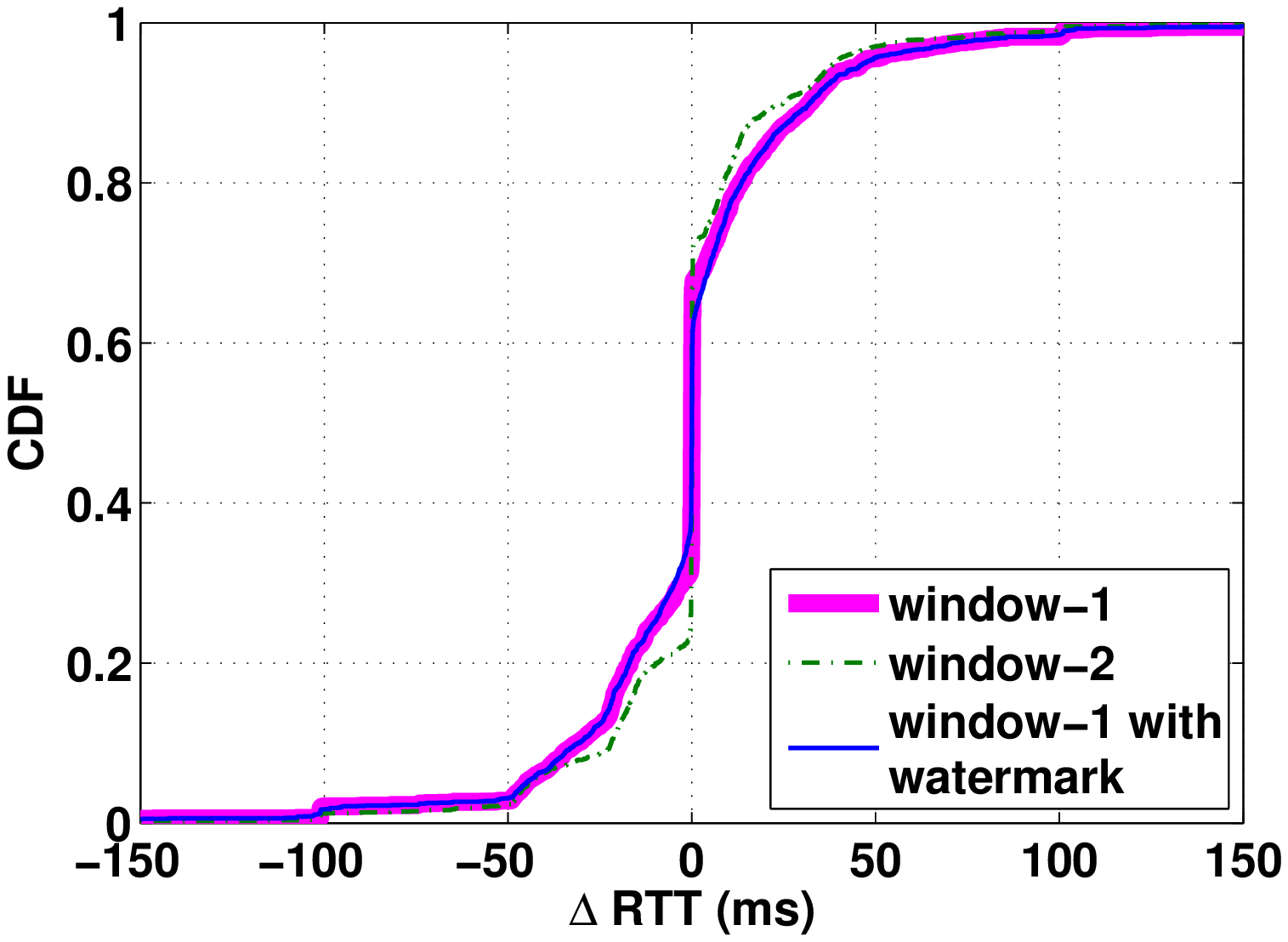}
   \label{fig:host2}
   }\subfigure[PlanetLab node: 128.138.207.54]{
  \includegraphics[width=0.32\columnwidth]{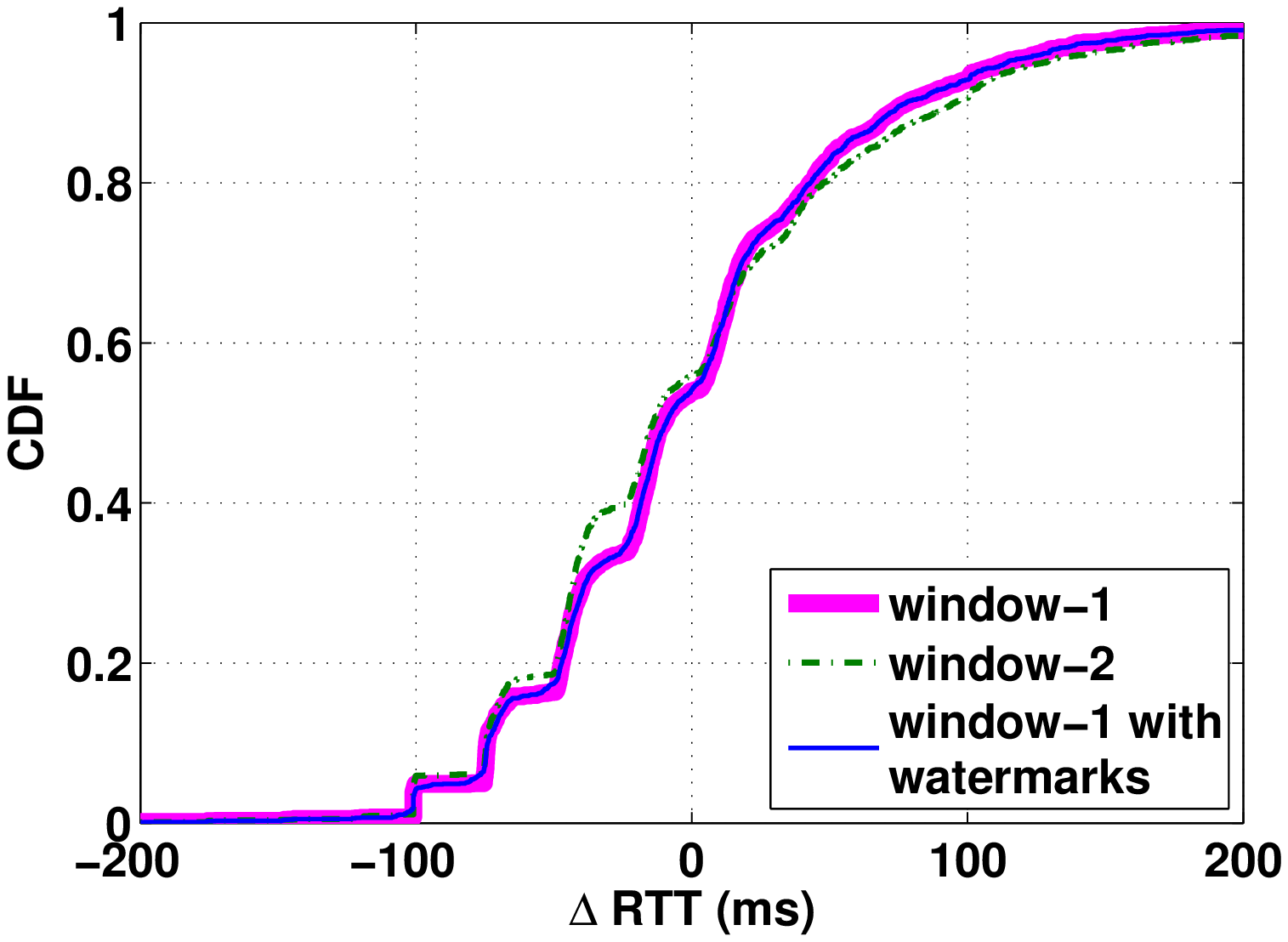}
   \label{fig:host3}
   }
   \subfigure[PlanetLab node: 192.12.33.100]{
   \includegraphics[width=0.32\columnwidth]{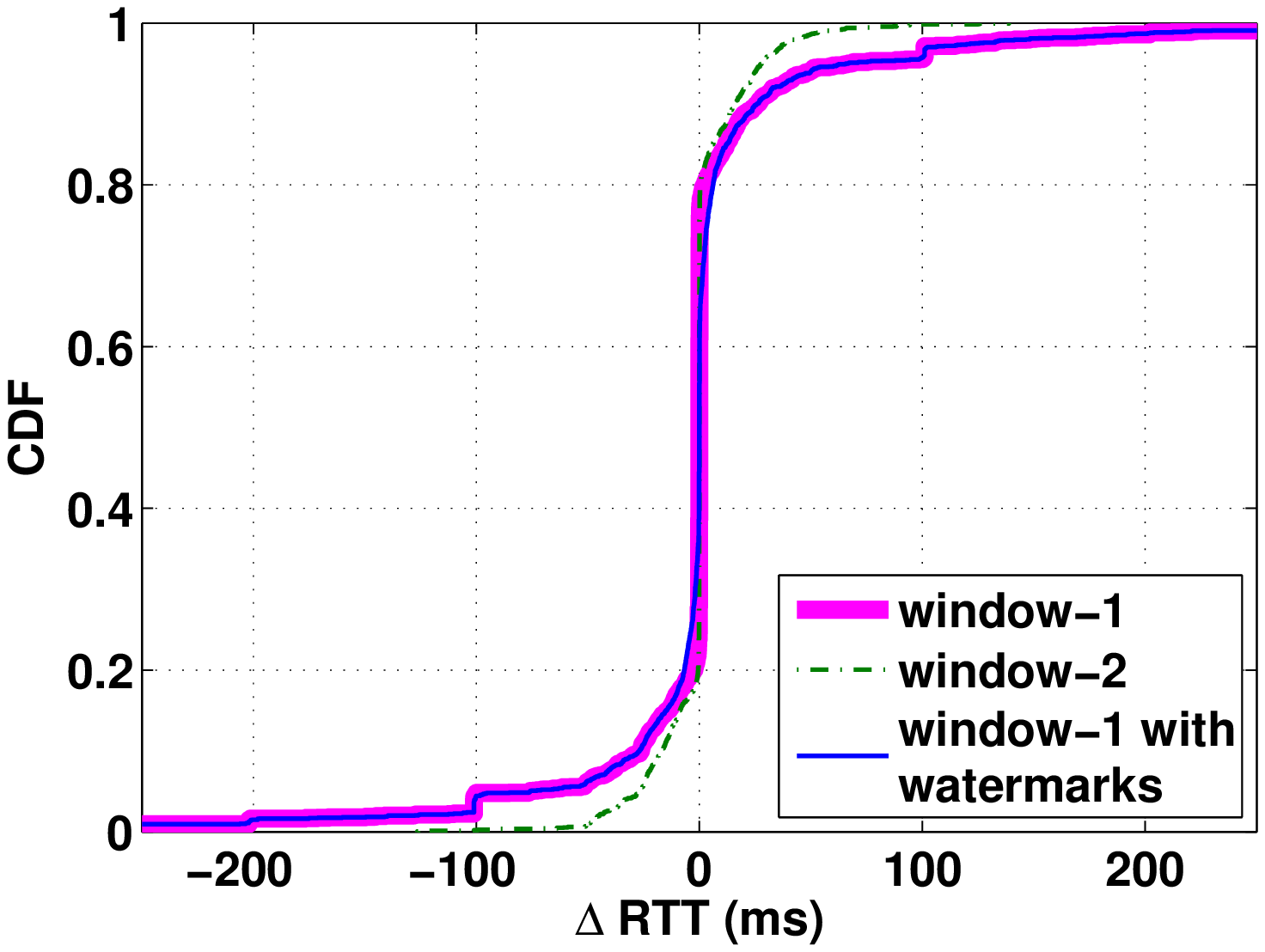}
   \label{fig:host4}
   }\subfigure[PlanetLab node: 206.117.37.6]{
   \includegraphics[width=0.32\columnwidth]{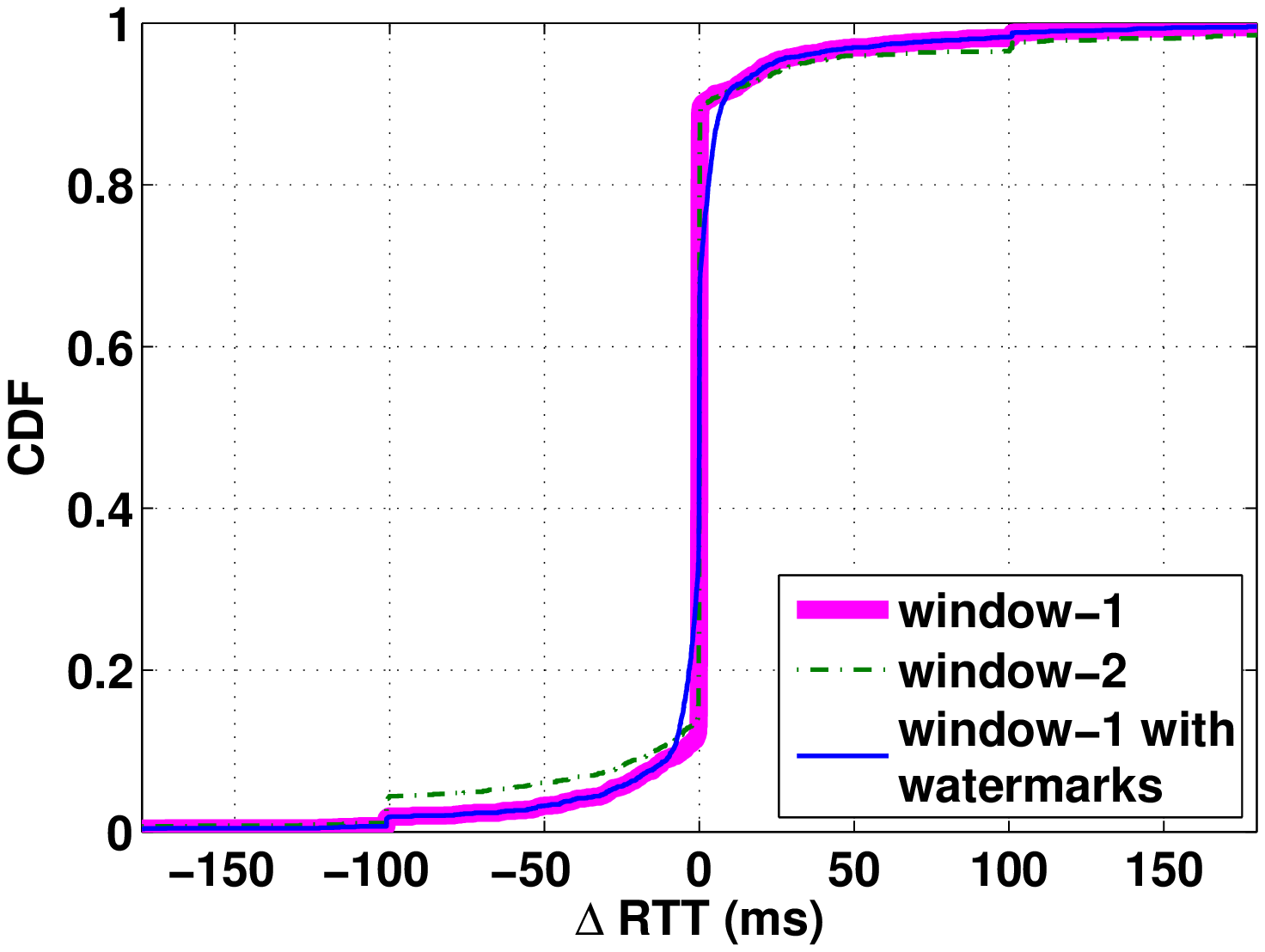}
   \label{fig:host5}
   }
   \caption{Comparisons of the $\Delta RTT$ distributions before and after adding watermarking.}  
    \label{fig:delta_rtt}
\end{figure}
Figure~\ref{fig:delta_rtt} depicts the CDFs of $\Delta RTT$ of Window-1, Window-1 with watermarks and Window-2.
We notice that except in Figure~\ref{fig:host1}, where the RTT jitter to the destination node was extremely low, the watermarked flow is not distinguishable from the unwatermarked flow. 
Our results indicate that BACKLIT only works in ``clean'' environments with negligible jitter, and the subtle watermark we inject remains invisible when moderate jitter exists.

To achieve simultaneous watermark robustness and invisibility, we embed a sparse watermark using the QIM embedding into flow IPDs. Modeling the network jitter, deletions, and insertions as a communication channel descried by a HMM, and employing an IDS decoder, we can reliably decoder the watermark. The QIM embedding meanwhile guarantees that watermark remains invisible to attackers.

\bibliographystyle{IEEETran}
\bibliography{sigproc}

\begin{thebibliography}{10}
\providecommand{\url}[1]{#1}
\csname url@samestyle\endcsname
\providecommand{\newblock}{\relax}
\providecommand{\bibinfo}[2]{#2}
\providecommand{\BIBentrySTDinterwordspacing}{\spaceskip=0pt\relax}
\providecommand{\BIBentryALTinterwordstretchfactor}{4}
\providecommand{\BIBentryALTinterwordspacing}{\spaceskip=\fontdimen2\font plus
\BIBentryALTinterwordstretchfactor\fontdimen3\font minus
  \fontdimen4\font\relax}
\providecommand{\BIBforeignlanguage}[2]{{%
\expandafter\ifx\csname l@#1\endcsname\relax
\typeout{** WARNING: IEEEtran.bst: No hyphenation pattern has been}%
\typeout{** loaded for the language `#1'. Using the pattern for}%
\typeout{** the default language instead.}%
\else
\language=\csname l@#1\endcsname
\fi
#2}}
\providecommand{\BIBdecl}{\relax}
\BIBdecl

\bibitem{xunICASSP12}
X.~Gong, M.~Rodrigues, and N.~Kiyavash, ``Invisible flow watermarks for
  channels with dependent substitution and deletion errors,'' in \emph{IEEE
  International Conf. Acoustics, Speech and Signal Processing (ICASSP)}, Kyoto,
  Japan, 2012, pp. 1773--1776.

\bibitem{Wang07}
X.~Wang, S.~Chen, and S.~Jajodia, ``Network flow watermarking attack on
  low-latency anonymous communication systems,'' in \emph{IEEE Symposium on
  Security and Privacy}, Oakland, CA, USA, 2007, pp. 116--130.

\bibitem{Staniford-Chen95}
S.~Staniford-Chen and L.~T. Heberlein, ``{Holding intruders accountable on the
  Internet},'' in \emph{IEEE Symposium on Security and Privacy}, Oakland, CA,
  USA, 1995, pp. 39--49.

\bibitem{Zhang00}
Y.~Zhang and V.~Paxson, ``Detecting stepping stones,'' in \emph{USENIX Security
  Symposium}, Denver, CO, USA, 2000, pp. 171--184.

\bibitem{Yoda00}
K.~Yoda and H.~Etoh, ``Finding a connection chain for tracing intruders,'' in
  \emph{Proc. 6th European Symposium on Research in Computer Security}, London,
  UK, 2000, pp. 191--205.

\bibitem{Wang02}
X.~Wang and D.~S. Reeves, ``{Robust correlation of encrypted attack traffic
  through stepping stones by manipulation of interpacket delays},'' in
  \emph{Proc. 10th ACM Conf. on Computer and Communications Security (CCS)},
  Washington, DC, USA, 2003, pp. 20--29.

\bibitem{Wang05}
X.~Wang, S.~Chen, and S.~Jajodia, ``Tracking anonymous peer-to-peer voip calls
  on the {Internet},'' in \emph{Proc. 12th ACM conference Computer and
  Communications Security (CCS)}, Alexandria, VA, USA, 2005.

\bibitem{Pyun07}
Y.~J. Pyun, Y.~H. Park, X.~Wang, D.~S. Reeves, and P.~Ning, ``Tracing traffic
  through intermediate hosts that repacketize flows,'' in \emph{26th IEEE
  International Conf. on Computer Communications (Infocom)}, Anchorage, AK,
  USA, 2007, pp. 634--642.

\bibitem{Houmansadr09}
A.~Houmansadr, N.~Kiyavash, and N.~Borisov, ``Rainbow: A robust and invisible
  non-blind watermark for network flows,'' in \emph{Proc. of 16th Network and
  Distributed System Security Symposium (NDSS)}, San Diego, CA, USA, 2009.

\bibitem{Yu07}
W.~Yu, X.~Fu, S.~Graham, D.~Xuan, and W.~Zhao, ``Dsss-based flow marking
  technique for invisible traceback,'' in \emph{IEEE Symposium on Security and
  Privacy}, Oakland, CA, USA, 2007, pp. 18--32.

\bibitem{Kiyavash08}
N.~Kiyavash, A.~Houmansadr, and N.~Borisov, ``Multi-flow attacks against
  network flow watermarking schemes,'' in \emph{Proc. 17th Conf. on USENIX
  Security Symposium}, San Jose, CA, USA, 2008, pp. 307--320.

\bibitem{Chen01}
B.~Chen and G.~W. Wornell, ``Quantization index modulation: A class of provably
  good methods for digital watermarking and information embedding,'' \emph{IEEE
  Trans. Inf. Theory}, vol.~47, no.~4, pp. 1423--1443, May 2001.

\bibitem{Massey51}
E.~J. Massey, ``The kolmogorov-smirnov test for goodness of fit,''
  \emph{Journal of the American Statistical Association}, vol.~46, no. 253, pp.
  68--78, 1951.

\bibitem{ssh}
\BIBentryALTinterwordspacing
RFC4252. The secure shell ({SSH}) authentication protocol. [Online]. Available:
  \url{http://tools.ietf.org/html/rfc4252}
\BIBentrySTDinterwordspacing

\bibitem{Yung02}
K.~H. Yung, ``Detecting long connection chains of interactive terminal
  sessions,'' in \emph{Proc. 5th International Conf. Recent Advances in
  Intrusion Detection(RAID)}, Zurich, Switzerland, 2002, pp. 1--16.

\bibitem{Ding11}
W.~Ding and S.-H. Huang, ``Detecting intruders using a long connection chain to
  connect to a host,'' in \emph{IEEE International Conf. Advanced Information
  Networking and Applications (AINA)}, Biopolis, Singapore, 2011, pp. 121--128.

\bibitem{Lippmann05}
R.~P. Lippmann, K.~W. Ingols, C.~Scott, K.~Piwowarski, K.~J. Kratkiewicz,
  M.~Artz, and R.~K. Cunningham, ``{Evaluating and strengthening enterprise
  network security using attack graphs},'' MIT Lincoln Lab., Tech. Rep.
  PR-IA-2, Aug. 12, 2005.

\bibitem{i2p}
\BIBentryALTinterwordspacing
{I2P} anonymous network. [Online]. Available: \url{http://www.i2p2.de/}
\BIBentrySTDinterwordspacing

\bibitem{freenet}
\BIBentryALTinterwordspacing
The {Freenet} project. [Online]. Available: \url{https://freenetproject.org/}
\BIBentrySTDinterwordspacing

\bibitem{gnunet}
\BIBentryALTinterwordspacing
{GNUnet}. [Online]. Available: \url{https://gnunet.org/}
\BIBentrySTDinterwordspacing

\bibitem{Cox}
I.~Cox, M.~Miller, and J.~Bloom, \emph{Digital Watermarking}.\hskip 1em plus
  0.5em minus 0.4em\relax Morgan Kaufmann Publisher, 2001.

\bibitem{Sergio}
S.~D. Servettot, C.~I. Podilchuks, and K.~Ramchandrant, ``Capacity issues in
  digital image watermarking,'' in \emph{International Conf. Image Processing
  (ICIP)}, Chicago, IL, USA, 1998, pp. 445--449.

\bibitem{Lin}
Z.~Lin and N.~Hoppers, ``New attacks on timing-based network flow watermarks,''
  in \emph{Proc. of the 21st USENIX Conf. on Security symposium}, Bellevue, WA,
  USA, 2012, pp. 20--35.

\bibitem{Luo}
X.~Luo, P.~Zhou, J.~Zhang, R.~Perdisci, W.~Lee, and R.~Chang, ``Exposing
  invisible timing-based traffic watermarks with backlit,'' in \emph{Proc. of
  the 27th Annual Computer Security Applications Conference}, Orlando, FL, USA,
  2011, pp. 197--206.

\bibitem{Davey00}
M.~C. Davey and D.~J.~C. MacKay, ``{Watermark codes: reliable communication
  over insertion/deletion channels},'' in \emph{Proc. IEEE International
  Symposium in Information Theory (ISIT)}, Sorrento, Italy, 2000, pp. 477--477.

\bibitem{Davey01}
M.~C. Davey and D.~J.~C. Mackay, ``Reliable communication over channels with
  insertions, deletions, and substitutions,'' \emph{IEEE Trans. Information
  Theory}, vol.~47, no.~2, pp. 687--698, Feb. 2001.

\bibitem{CAIDA}
\BIBentryALTinterwordspacing
{The Cooperative Association for Internet Data Analysis}. [Online]. Available:
  \url{http://www.caida.org/}
\BIBentrySTDinterwordspacing

\bibitem{Swirl}
A.~Houmansadr and N.~Borisov, ``Swirl: A scalable watermark to detect
  correlated network flows,'' in \emph{Proc. of 16th Network and Distributed
  System Security Symposium (NDSS)}, San Diego, CA, USA, 2011.

\bibitem{Planetlab}
\BIBentryALTinterwordspacing
{PlanetLab: An open platform for developing, deploying, and accessing
  planetary-scale services}. [Online]. Available:
  \url{http://www.planet-lab.org/}
\BIBentrySTDinterwordspacing

\end{thebibliography}

\end{document}